\documentclass[useAMS,usenatbib]{mn2e}
\usepackage{amsmath,epsf}

\def\lesssim{\mathrel{\hbox{\rlap{\hbox{\lower4pt\hbox{$\sim$}}}\hbox{$<$}}}}
\def\gtrsim{\mathrel{\hbox{\rlap{\hbox{\lower4pt\hbox{$\sim$}}}\hbox{$>$}}}}
\newcommand{\ltaraw}{$\; \buildrel < \over \sim \;$}
\newcommand{\lta}{\lower.5ex\hbox{\ltaraw}}
\newcommand{\gtaraw}{$\; \buildrel > \over \sim \;$}
\newcommand{\gta}{\lower.5ex\hbox{\gtaraw}}

\newcommand{\ie}{{\it i.e.~}}
\newcommand{\eg}{{\it e.g.~}}

\newcommand{\etal}{{\it et.~al.~}}

\newcommand{\Chandra}{\emph{Chandra}}
\newcommand{\XMM}{XMM{}}

\def\araa{ARA\&A}%
\def\apj{ApJ}%
\def\apjl{ApJ}%
\def\apjs{ApJS}%
\def\aap{A\&A}%
\def\mnras{MNRAS}%
\newcommand{\trelax}{$t_{relax}$}

\newcommand{\taccrete}{$t_{accrete}$}
\newcommand{\tclosest}{$t_{closest}$}
\newcommand{\tapo}{$t_{apo}$}

\newcommand{\kpc}{\,\mbox{kpc}}

\newcommand{\Mpc}{\,\mbox{Mpc}}

\newcommand{\eV}{\,\mbox{eV}}

\newcommand{\cm}{\,\mbox{cm}}
\newcommand{\Gyr}{\,\mbox{Gyr}}

\newcommand{\keVcmsq}{\,\mbox{keVcm$^{2}$}}
\newcommand{\ks}{\,\mbox{ks}}
\newcommand{\K}{\,\mbox{K}}

\newcommand{\kms}{\, \mbox{km}\,\mbox{s$^{-1}$}}
\newcommand{\keV}{\,\mbox{keV}}
\newcommand{\msun}{\,M_{\sun}}

\newcommand{\Gyrs}{\,\mbox{Gyr}}
\newcommand{\Myrs}{\,\mbox{Myr}}

\newcommand{\tsim}{\sim\!}

\newcommand{\URL}{http://astro.phys.uvic.ca/$\sim$babul/Merger\_PaperI}

\title[The impact of mergers on relaxed X-ray clusters I]
  {The impact of mergers on relaxed X-ray clusters \\ 
   I. Dynamical evolution and emergent transient structures}

\author[Gregory B. Poole \etal]{Gregory B. Poole$^{1}$\thanks{E-mail: gbpoole@uvic.ca}, Mark A. Fardal$^2$, Arif Babul$^1$, Ian G. McCarthy$^3$, \newauthor
  Thomas Quinn$^4$, and James Wadsley$^5$\\
  $^{1}$Dept. of Physics \& Astronomy, University of Victoria, 
  Elliott Building, 3800 Finnerty Rd., Victoria, BC, V8P 1A1, Canada\\
  $^2$Dept.\ of Astronomy, University of Massachusetts, 
      Amherst, MA, 01003, USA\\
  $^3$Department of Physics, University of Durham, 
      South Road, Durham DH1 3LE, UK\\
  $^4$Dept.\ of Astronomy, University of Washington, Seattle, WA 98195\\ 
  $^5$Dept.\ of Physics \& Astronomy, McMaster University, Hamilton, Ontario, L88 4M1, Canada
  }
\date{draft version \today}
\pagerange{\pageref{firstpage}--\pageref{lastpage}} \pubyear{2006}

\def\LaTeX{L\kern-.36em\raise.3ex\hbox{a}\kern-.15em
    T\kern-.1667em\lower.7ex\hbox{E}\kern-.125emX}

\begin{document}

\label{firstpage}

\maketitle

\begin{abstract}
We report on the analysis of a suite of SPH simulations (incorporating cooling and star formation)  of mergers involving idealised X-ray clusters
whose initial conditions  resemble relaxed clusters with cool compact cores observed by \Chandra\ and \XMM.  The simulations sample the most
interesting, theoretically plausible, range of impact parameters and progenitor mass ratios.  We find that all mergers evolve via a common
progression.  We illustrate this progression in the projected gas density, X-ray surface brightness, SZ, temperature, and gas entropy maps.
Several different classes of transient ``cold front''-like features can arise over the course of a merger. Each class is distinguished by a
distinct morphological signature and physical cause.  We find that all of these classes are present in \Chandra\ and \XMM\ observations of 
merging systems and propose a naming scheme for these features: ``comet-like'' tails, bridges, plumes, streams and edges.  In none of the cases 
considered do the initial cool compact cores of the primary and the secondary 
get destroyed during the course of the mergers. Instead, the two remnant cores eventually combine to form a new core that, depending on the final 
mass of the remnant, can have a greater cooling 
efficiency than either of its progenitors.   We quantify the evolving morphology of our mergers using centroid variance, power ratios and offset
between the X-ray and the projected mass maps.  We find that the centroid variance best captures the dynamical state of the cluster.  
It also provides an excellent indicator of how far the system is from virial and hydrostatic equilibrium.  Placing the system at
$z=0.1$, we find that all easily identified observable traces of the secondary disappear from a simulated $50 \ks$ \Chandra\ image following the
second pericentric passage.  The system, however, takes an additional $\sim 2$\Gyrs~to relax and virialize.  Observationally, the only reliable
indicator of a system in this state is the smoothness of its the X-ray surface brightness isophotes, not temperature fluctuations.  Temperature
fluctuations at the level of $\Delta T/T\sim 20\%$, can persist in the final systems well past the point of virialization, suggesting that 
that the existence of temperature fluctuations, in and of themselves, do not necessarily indicate a disturbed or unrelaxed system. 

\end{abstract}


\section{Introduction}\label{sec-intro} 
The utility of galaxy clusters as tools for studying cosmology and the history of structure formation is well established.  Furthermore, sizable samples of clusters extending to high redshift will soon be available from wide field X-ray and Sunyaev-Zel'dovich (SZ) surveys, greatly expanding their importance and ensuring that they will feature prominently in these pursuits throughout the foreseeable future.

Often implicit in such studies is the assumption that clusters are in equilibrium.  However, galaxy clusters are dynamically young systems and are thus prone to regular disruptions by mergers \citep{CohnWhite05}.  Recent \Chandra\ and \XMM\ observations support this.  High resolution spectroscopic imaging has provided dramatic evidence of ongoing mergers in several systems.  As a result, we now know that the intracluster medium (ICM) of merging clusters can express a variety of complicated transient features including cold fronts \citep{Markevitchetal00}, shock fronts \citep{MV01} and the ``sloshing'' of cool cores in their dark matter potentials \citep{Markevitchetal01,DupkeWhite03}. 

Such observations have stimulated several theoretical and computational studies devoted to understanding the impact of mergers on clusters.  \citet{Bialeketal02}, for instance, have shown that cold fronts can be produced through the disruption of cold cores during mergers while \citet{Motl04} have shown that a significant fraction of a merging cold core can avoid shock heating, contributing to the assembly of the remnant's cold core.  Furthermore, \citet{RTK04} have shown that smooth accretion and mergers lead to different evolutions in a system's X-ray luminosity and temperature.  Combined with the results of \citet{OHaraetal05} who suggest that all clusters may exhibit departures from equilibrium due to the lingering effects of their past merger activity, it becomes clear that we must understand the impact of merging in accounting for the observed properties of clusters; even the seemingly relaxed ones.

Determining the detailed effects of a single merger on a cluster evolving in a cosmological environment can be greatly complicated by the presence of extant substructure, subsequent mergers and smooth accretion.  Simulations of idealised two-body cluster mergers provide an alternative approach and, given the fact that most clusters obtain much of their mass through significant mergers \citep{CohnWhite05}, are not a significant departure from cosmological simulations.  They allow control of the initial structure and orbits of the interacting systems and facilitate the investigation of not only the obvious short-term effects of the event, but more subtle long-term effects as well.  They also simplify the interpretation of specific merger observations which typically involve only two dominant components.

Several authors have explored cluster mergers in this way \citep[see][for pioneering studies]{Roettigeretal93,SchindlerMueller93,Pearceetal94,Roettigeretal97}.  Amoung recent studies, \citet{Gomezetal} have constructed a suite of idealised merging systems of 4:1 and 16:1 mass ratios to study the effects of mergers on the stability of central cool cores.  The initial conditions for their clusters were chosen to have gas core radii of $r_c\sim250$\kpc\ and the dark matter in their simulations was distributed according to a King model.  \citet[][RS01 hereafter]{RickerandSarazin01} have studied mergers between idealised systems with cuspy dark matter profiles.  Since the focus of their study was on the luminosity and temperature ``boosts'' which a merger system experiences upon impact of the cluster cores, they did not include the effects of cooling in their simulations.  Lastly, \citet{RT02} have analysed a set of simulations incorporating cooling and star formation for idealised merging systems constructed with cuspy dark matter profiles.  Their initial conditions primarily examine systems with long cooling times and core radii of $r_c\sim100$\kpc, although three cases represent systems with short cooling times and small core radii ($r_c\sim50$\kpc).

However, it is now known that clusters have cuspy dark matter profiles and that $70-90$\% of observed systems possess compact cores \citep[$r_c \sim 50$\kpc,][]{Peresetal98,Edgeetal92} with cooling times short enough to be of significant importance to their structure.  Hence, a comprehensive study of cluster mergers utilising initial conditions faithful to our contemporary understanding of a typical cluster's structure while including the important effects of cooling has not been performed.  Given the wide ranging applications of cluster studies, the importance of mergers to understanding the structure of the ICM and the wealth of detailed cluster merger observations requiring interpretation, a thorough revisit of the issues introduced by cluster mergers would be of great utility.

In this paper and the series to follow, we will present an analysis of a suite of SPH simulations of isolated two-body mergers between idealised clusters constructed to possess realistic compact cool cores.  With this study we seek to provide a conceptual foundation for subsequent studies to be conducted in a full cosmological context.  Our approach utilises insights provided by recent high resolution observations of clusters to motivate the initial conditions for a set of nine hydrodynamic simulations (incorporating the effects of cooling and star formation) involving a representative range of mass ratios and impact parameters.  In this article we explain our approach to constructing idealised merger simulations and present their general dynamical progression, eventual relaxation and describe the transient phenomena which manifest during the event.  Since many of the phenomena we study have low surface brightnesses, we utilise synthetic observations to delineate realistically observable phenomena from those which are not.  In subsequent papers we will examine in detail several specific issues which we can only touch upon in the present work.

In Section \ref{sec-method} we describe our methods for initialising merging systems and placing them on realistic orbits as well as the details of our numerical methods.  In Section \ref{sec-analysis} we qualitatively describe the  evolution of our cluster mergers through a generic sequence of states and introduce the transient structures formed in the process.  In Section \ref{analysis-disturbed} we examine the evolution of several measures of our mergers' apparent degree of disruption to determine when our systems would appear undisturbed under reasonable observational circumstances.  In Section \ref{analysis-relaxation} we assess the degree to which our apparently relaxed merger remnants are formally so.  In Section \ref{analysis-transients} we examine in further detail the different transient structures which form during a merger, describing their properties and the processes which drive their creation.  Finally we summarise our study in Section \ref{sec-summary}.

In all cases our assumed cosmology will be ($\Omega_M$,$\Omega_\Lambda$)=($0.3$,$0.7$) with $H_0$=$75$~km/s/Mpc and $\Omega_b = 0.02 h^{-2}$.

\section{Simulations}\label{sec-method}
\begin{table*}
\begin{minipage}{155mm}
\caption{Initial separations ($r$), radial and transverse velocities ($v_r$ and $v_t$ respectively) for our simulation orbits and measures of the final remnant's global structure (the spin parameter $\lambda_{200}$ and triaxial ratios $q_{200}$ and $s_{200}$, all measured within $R_{200}$).  At the start of our simulations, $t=0$; when the secondary's centre of mass passes the virial radii $R_{100}$ or $R_{200}$ of the primary, $t=t'_o$ or $t=t_o$ respectively.\label{table-initial_conditions}}
\begin{tabular}{cccccccccc}
\hline
$M_p$:$M_s$         &
$v_t(t'_o)/v_c(t'_o)$ & 
$t_o [Gyrs]$        & 
$r(0) [Mpc]$        & 
$v_{t}(0) [km/s]$   & 
$v_{r}(0) [km/s]$   & 
$r_{min} [kpc]$     &
$\lambda_{200}$     &
$q_{200}$           &
$s_{200}$           \\         
\hline
1:1   &  0.00  & 3.4 &  6.54  &    0  &  $-721$  &    5   & 0.0    & 0.67 & 0.62 \\
1:1   &  0.15  & 3.4 &  6.54  & -111  &  $-708$  &  110   & 0.021  & 0.70 & 0.65 \\
1:1   &  0.40  & 3.2 &  6.54  & -315  &  $-925$  &  390   & 0.059  & 0.68 & 0.65 \\
3:1   &  0.00  & 3.4 &  5.53  &    0  &  $-574$  &   12   & 0.0    & 0.76 & 0.73 \\
3:1   &  0.15  & 3.4 &  5.53  &  -87  &  $-587$  &  103   & 0.014  & 0.80 & 0.75 \\
3:1   &  0.40  & 2.6 &  5.53  & -271  &  $-908$  &  360   & 0.042  & 0.78 & 0.76 \\
10:1  &  0.00  & 2.2 &  4.81  &    0  &  $-905$  &   10   & 0.0    & 0.96 & 0.94 \\
10:1  &  0.15  & 2.4 &  4.81  & -106  &  $-872$  &  130   & 0.007  & 0.96 & 0.92 \\
10:1  &  0.40  & 2.2 &  4.81  & -284  &  $-950$  &  419   & 0.018  & 0.97 & 0.93 \\
\hline
\end{tabular}
\end{minipage}
\end{table*}

We run our simulations with GASOLINE \citep{wadsley04}, a versatile parallel SPH tree code with multi-stepping.  We include the effects of radiative cooling, star formation and minimal feedback from supernovae in our simulations but the effects of feedback from active galactic nuclei (AGN) are omitted.  It is important to note that we do not include the effects of magnetic fields, pressure generated from cosmic rays or conduction and that our choice of algorithm is not well suited to modelling the effects of turbulence.  The relevance of each of these processes is not well established (theoretically or observationally) and we can only provide the caveat that some of our results are susceptible to change once we have the resources to properly model them \citep[although, see][for references suggesting that neither conduction nor turbulence are relevant in cluster cores]{Markevitchetal03,Fabianetal03}.

In this section we describe how our clusters are initialised, how we set up pairs of clusters to form a merging system, the numerical methods and code parameters of our runs.

\subsection{Cluster initial conditions}\label{numerics-cluster_initial}

The structure of a merger remnant is likely to depend sensitively on the initial structure of the interacting systems \citep{Gomezetal} making it important to implement realistic initial conditions for the gas and dark matter properties of our clusters.  We follow the analytic prescription of \citet{BBLP} and \citet{MBBPH} to produce systems which conform with recent theoretical and observational insights into cluster structure.

 The dark matter density profiles of our systems follow an NFW-like form \citep{NFW96,Mooreetal98} given by
\begin{center}
\begin{equation}\label{eqn-NFW}
\rho_{DM}(r)=\frac{\rho_s}{(r/r_s)^{\beta}\left(1+(r/r_s)\right)^{3-\beta}}
\end{equation}
\end{center}

\begin{figure*}
\begin{minipage}{175mm}
\begin{center}
\epsfysize=95mm \epsfbox{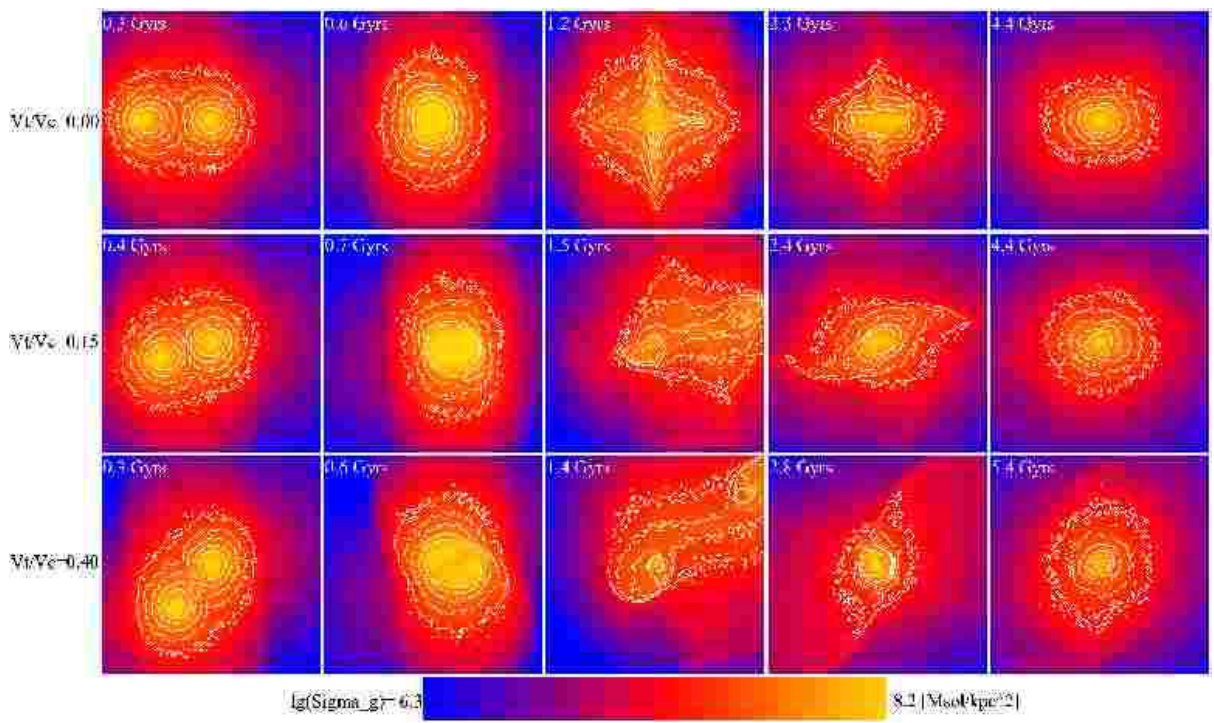}
\caption{Maps of the gas surface density for our 1:1 simulations, projected along an axis normal to the plane of the orbit ($3$\Mpc\ on a side).  White contours are X-ray surface brightness isophotes from simulated 50\ks\ $z=0.1$ \Chandra\ observations ($3\times$ to $48\times$ the background stepping by factors of $2$).  Times depicted are \tclosest$-200$\Myrs\ (where \tclosest is the time of first pericentric passage), \tclosest$+100$\Myrs, \tapo (where \tapo is the time of first apocentric passage), \taccrete$+100$\Myrs~ (where \taccrete is the time of the second and usually the final pericentric passage) and \trelax, the time when system appears relaxed. All times are measured relative to $t_o$ (the time when the secondary's centre of mass first crosses $R_{200}$ of the primary)$^\dagger$.}
\label{fig-1to1_Sigmagas}
\end{center}
\end{minipage}
\end{figure*}

\noindent with the central asymptotic logarithmic slope chosen to be $\beta=1.4$ and $r_s$ selected to yield a concentration $c=R_{200}/r_s=2.6$ (we will use $R_{\Delta}$ throughout to indicate the radius within which the mean density of the system is $\Delta$ times the critical density, $\rho_c=3 H_0^2/8 \pi G$).  This dark matter distribution would yield $c=4.5$ if fit by Eqn. \ref{eqn-NFW} with $\beta=1$ and is consistent with results from cosmological N-body simulations \citep{Ekeetal01}.

\renewcommand{\thefootnote}{\fnsymbol{footnote}}\footnotetext[2]{For colour plots, please see the electronic version or \URL}\renewcommand{\thefootnote}{\arabic{footnote}}

The initial density and temperature profiles of the clusters are set by requiring that (1) the gas be in hydrostatic potential within the halo, (2) the ratio of gas mass to dark matter mass within the virial radius be $\Omega_b/(\Omega_m - \Omega_b)$, and (3) the initial gas entropy$^6$\footnotetext[6]{We use the standard proxy for entropy given by $S\equiv kT/n_e^{2/3}$ with $n_e$ and $T$ representing the electron density and temperature of the gas.} scale as  $S(r) \propto r^{1.1}$ over the bulk of the cluster body. 

Our choice for the initial gas distribution is motivated by the fact that we are interested in studying ``typical'' systems which correspond to the most likely pre-interaction configuration for merging clusters.  Estimates in the literature suggest that the majority of observed clusters ($70$--$90$\%) possess dense, cool, compact cores with short cooling times \citep{Peresetal98}.  Recent \Chandra\ observations reveal that these systems exhibit power-law entropy profiles of the form $S(r) \propto r^{1.0-1.3}$, beyond the central $10$\kpc,  over a wide range of mass scales \citep{Donahueetal05}.  We note that our choice for the form of the initial density profile conforms not only with these observations but also with the results of high resolution cluster simulations \citep{Lewisetal00,Voitetal03}.  

We normalize the entropy profiles such that the temperature of the ICM at $R_{vir}$ is half the virial temperature.  This also approximately matches cosmological simulations \citep{Lewisetal00,Lokenetal02} and observations \citep{DeGrandiandMolendi02}. Finally, for practical purposes, we start our systems with small, constant, low-entropy ($10$\keVcmsq) cores.  The cores, however, 
evolve to equilibrium power-law distributions quickly ($\sim 0.5$\Gyrs) and  well before the clusters begin interacting significantly.

Particle realizations are initialised using the ZENO package of J. Barnes.  Gas particles are distributed spherically and assigned temperatures according to the model described above.  The dark matter velocity distribution is taken to be isotropic and the distribution function of particle energies calculated by solving the Abel integral equation \citep{BandT,kazantzidis04}.  Previous cluster merger studies have been initialised with Gaussian distributions resulting in systems which subsequently evolve significantly from their initial conditions (RS01).  Our approach initialises the system very close to equilibrium, ensuring that our interacting systems accurately maintain our desired initial conditions until they collide.

Negative values for the dark matter energy distribution function are avoided by truncating our clusters (dark matter and gas) smoothly beyond the virial radius ($R_{vir}$) with the function $\rho(r) \propto (r/R_{vir})^\gamma \exp(-[(r-R_{vir})/R_{sm}]^3)$, where $\gamma$ is set to ensure a continuous first derivative for $\rho(r)$.  Beyond the point where the gas pressure reaches that of the intergalactic medium (IGM) (we choose $P_{IGM} = (\Omega_b \rho_c \, k T_{IGM}) / (\mu m_H)$ with $T_{IGM} = 3 \times 10^5 \K$), we surround our systems with a dynamically negligible uniform gaseous medium of pressure $P_{IGM}$.  This is done mainly to confine small numbers of high-velocity gas particles occasionally produced during our simulations.  Without this external medium, these particles would adiabatically expand to unrealistically low densities and temperatures and pose numerical difficulties for the code.

\begin{figure*}
\begin{minipage}{175mm}
\begin{center}
\epsfysize=95mm \epsfbox{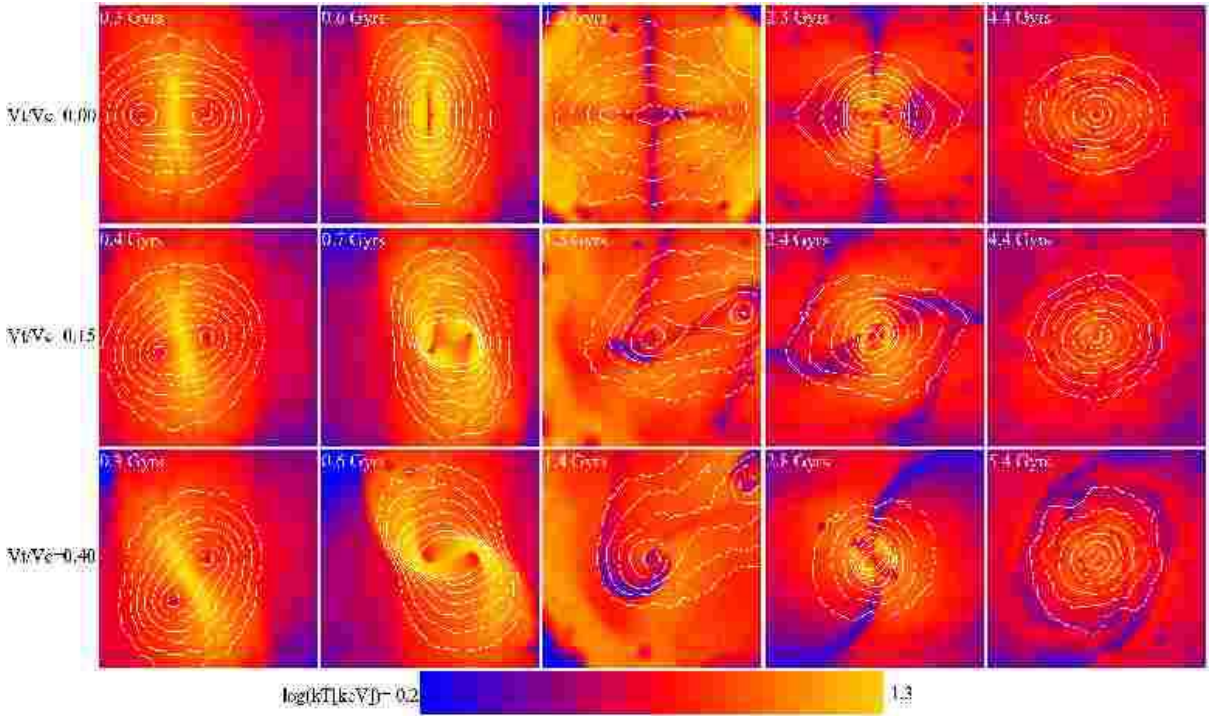}
\caption{Maps of the projected ``spectroscopic-like'' temperature \citep{Mazzottaetal04} maps for a $0.5$\Mpc\ thick slice ($3$\Mpc\ on a side) passing through the centre of our $1:1$ simulations.  Contours depict the SZ effect (for the entire simulation volume along the line of sight) and represent $log(y)=-5.5$ to $-4$ in increments of $0.25$$^\dagger$.  The times represented are the same as Fig. \ref{fig-1to1_Sigmagas}.}
\label{fig-1to1_Tsl}
\end{center}
\end{minipage}
\end{figure*}
\begin{figure*}
\begin{minipage}{175mm}
\begin{center}
\epsfysize=95mm \epsfbox{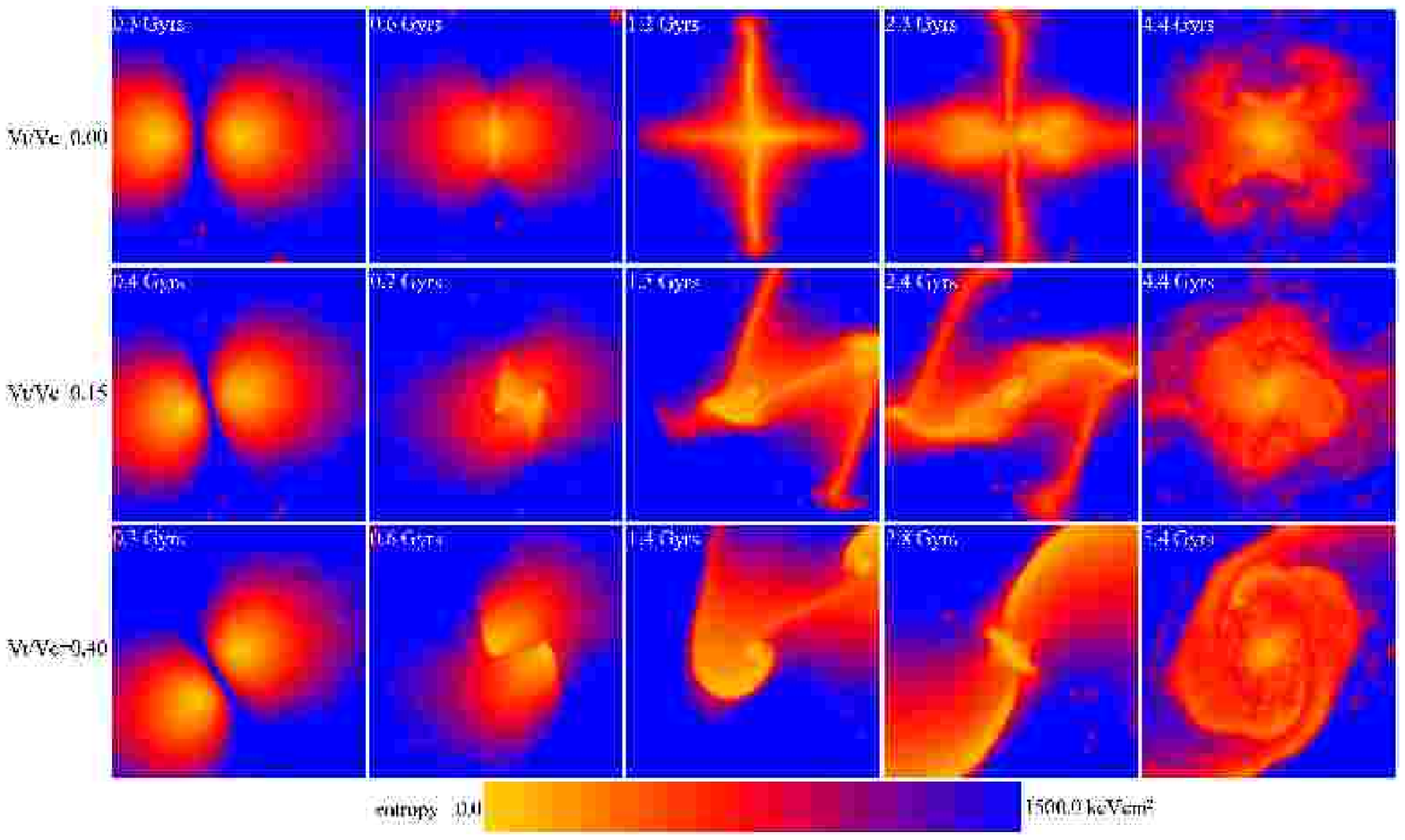}
\caption{Entropy maps for a $0.5$\Mpc\ thick slice ($3$\Mpc\ on a side) through the centres of our $1:1$ simulations$^\dagger$.  The times represented are the same as Fig. \ref{fig-1to1_Sigmagas}.}
\label{fig-1to1_entropy}
\end{center}
\end{minipage}
\end{figure*}

In the present study we examine mergers between systems set to have mass ratios of 1:1, 3:1 and 10:1.  In what follows we shall distinguish the two interacting clusters by identifying the most massive $10^{15} M_\odot$ system as the primary system (with mass $M_p$) and the smaller incident system (with mass $M_s$) as the secondary.  For our equal mass mergers, we arbitrarily choose one system as the primary with no effect on our results due to the symmetry in such cases.  All positions and velocities are measured relative to the centre of mass of the primary system.  

The virial mass of primary system is set to $10^{15} M_\odot$ in all cases and the secondary systems have virial masses of $10^{15} \msun$, $10^{14.5} \msun$, and $10^{14} \msun$.  These values are larger than $M_{200}$, which takes the values $7.5 \times 10^{14} \msun$, $2.4 \times 10^{14} \msun$, and $7.8 \times 10^{13} \msun$ for the three systems.  We shall, however, refer to the systems by their virial mass rather than $M_{200}$.
The circular velocities of the halos at $R_{200}$ ($V_{200}$) are $1340 \kms$, $920 \kms$, and $630 \kms$ respectively.

\subsection{Initial kinematics}

The problem of initialising the kinematics of idealised merging clusters has been faced by several authors \citep[\eg][RS01]{RT02}.  Whereas previous studies have initialised orbits based on analytic arguments, our initial conditions are motivated by analysis of the orbital properties of substructure in high-resolution cosmological dark matter simulations.  

\begin{figure*}
\begin{minipage}{175mm}
\begin{center}
\epsfysize=95mm \epsfbox{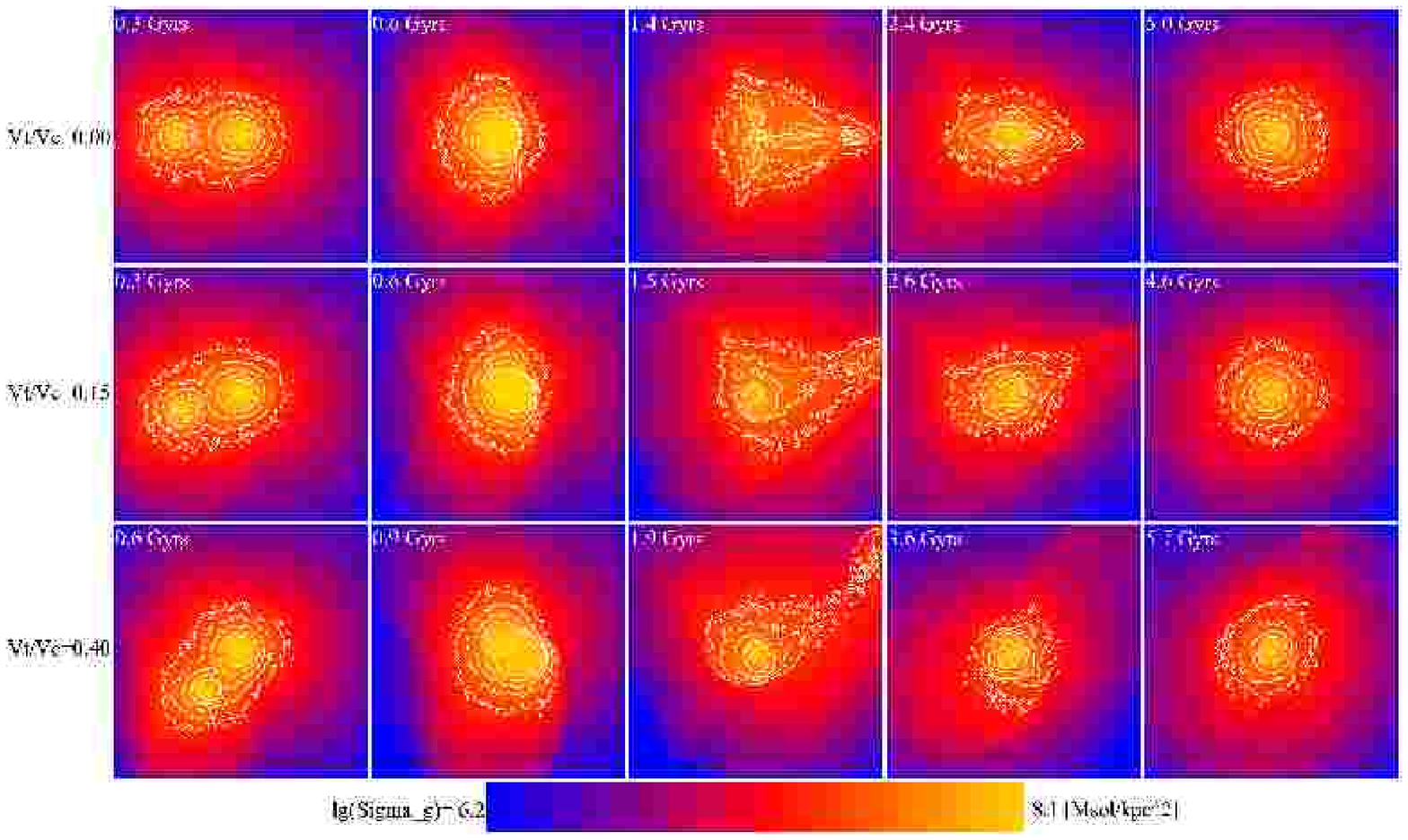}
\caption{Maps of the gas surface density for our 3:1 simulations, projected along an axis normal to the plane of the orbit ($3$\Mpc\ on a side).  White contours are X-ray surface brightness isophotes from simulated 50\ks\ $z=0.1$ \Chandra\ observations ($3\times$ to $48\times$ the background stepping by factors of $2$).  Times depicted are \tclosest$-200$\Myrs, \tclosest$+100$\Myrs, \tapo, \taccrete$+100$\Myrs~and \trelax. See Figure 1 caption or Table 2 for a brief definition of the timescales.  All times are measured relative to the $t_o$ when the secondary's centre of mass crosses the virial radius of the primary for the first time$^\dagger$.}
\label{fig-3to1_Sigmagas}
\end{center}
\end{minipage}
\end{figure*}

We construct orbits for these systems which produce specified radial and tangential velocities for the secondary system ($v_r$ and $v_t$ respectively) when its centre of mass reaches the virial radius of the primary ($R_{vir}$).  For each of the three mass ratios we study, we examine three orbits selected to produce a typical value of $v_r(R_{vir})$ and to cover a significant range of the transverse velocity $v_t(R_{vir})$ giving rise to mergers found in cosmological dark matter simulations.  

In an N-body study of haloes merging onto rich clusters, \citet{tormen97} find that the average velocity of an accreting secondary system is $v(R_{vir})=(1.1\pm0.1)V_c(R_{vir})$, where $V_{c}$ is the primary system's circular velocity.  In another study, \citet{vitvitska02} \citep[see also][]{Benson05} examine the distribution of secondary subhalo velocities moving within the virial radius of massive haloes.  They find that the total velocities of accreting systems (\ie those with negative radial velocities in the range $r=0.8$--$1.2R_{vir}$) relative to the primary are distributed normally.  They report mean infall velocities of $\langle v \rangle \approx V_c$ (independent of mass), consistent with \citet{tormen97}.  They also find that tangential 2-dimensional rms velocities decrease with the secondary mass, ranging from $\sigma_\perp \approx 0.4 V_c$ for 1:1 mergers to $\sigma_\perp \approx 0.67 V_c$ for 10:1 mergers.  It is important to note that these statistics likely include systems making secondary or tertiary encounters with their primary.  The orbits of such systems have likely been rendered more isotropic, making these tangential dispersions upper limits for substructure accreting for the first time.  To produce systems which will merge fully within a reasonable amount of time, we have chosen to examine three values of $v_t/V_c$ which roughly span the lower half of the \citet{Vitvitskaetal02} distribution.  Specifically, we construct orbits to produce $v_r(R_{vir})=1.2V_c(R_{vir})$ (slightly higher but consistent with the mean velocities cited above) and $v_t/V_c = 0$, $0.15$, and $0.4$.  Unlike in the rest of our study, we define the virial radius here by an overdensity $\Delta = 100$, corresponding to the threshold used by \citet{Vitvitskaetal02}.  Doing this for mass ratios of 1:1, 3:1 and 10:1 produces nine simulations which constitute the basis of the analysis we present in this study.  

\begin{figure*}
\begin{minipage}{175mm}
\begin{center}
\epsfysize=95mm \epsfbox{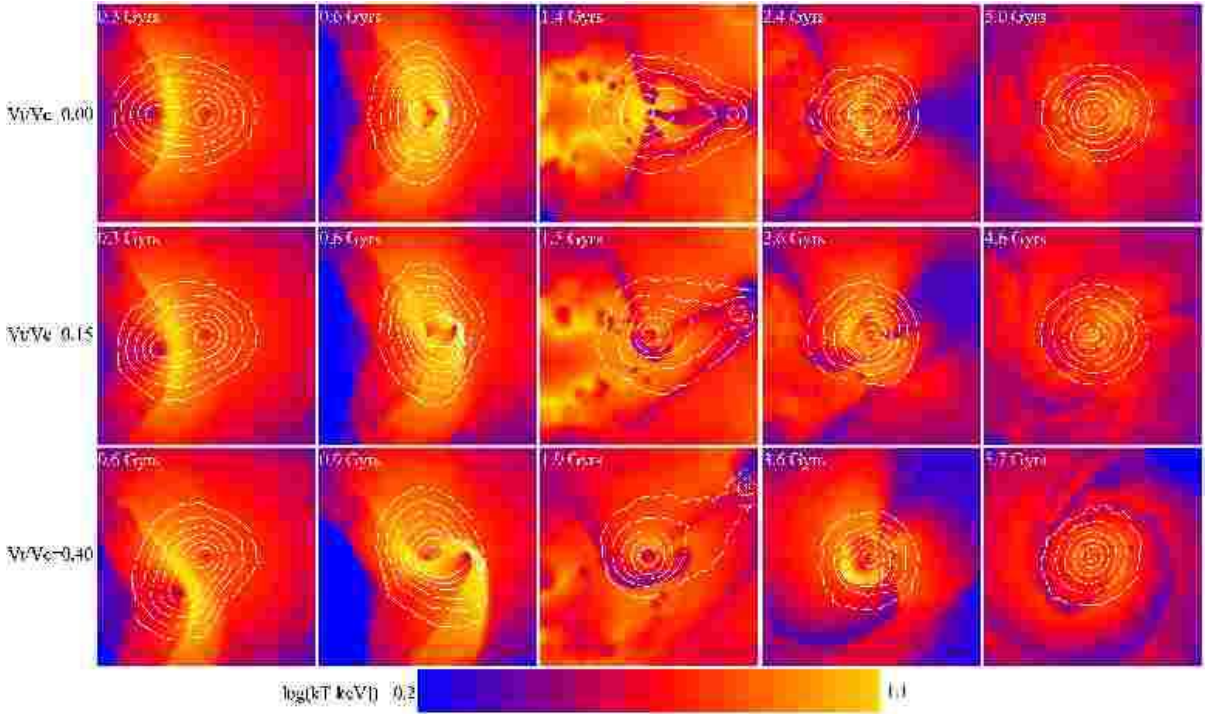}
\caption{Maps of the projected ``spectroscopic-like'' temperature \citep{Mazzottaetal04} maps for a $0.5$\Mpc\ thick slice ($3$\Mpc\ on a side) passing through the centre of our $3:1$ simulations.  Contours depict the SZ effect (for the entire simulation volume along the line of sight) and represent $log(y)=-5.5$ to $-4$ in increments of $0.25$$^\dagger$.  The times represented are the same as Fig. \ref{fig-3to1_Sigmagas}.}
\label{fig-3to1_Tsl}
\end{center}
\end{minipage}
\end{figure*}
\begin{figure*}
\begin{minipage}{175mm}
\begin{center}
\epsfysize=95mm \epsfbox{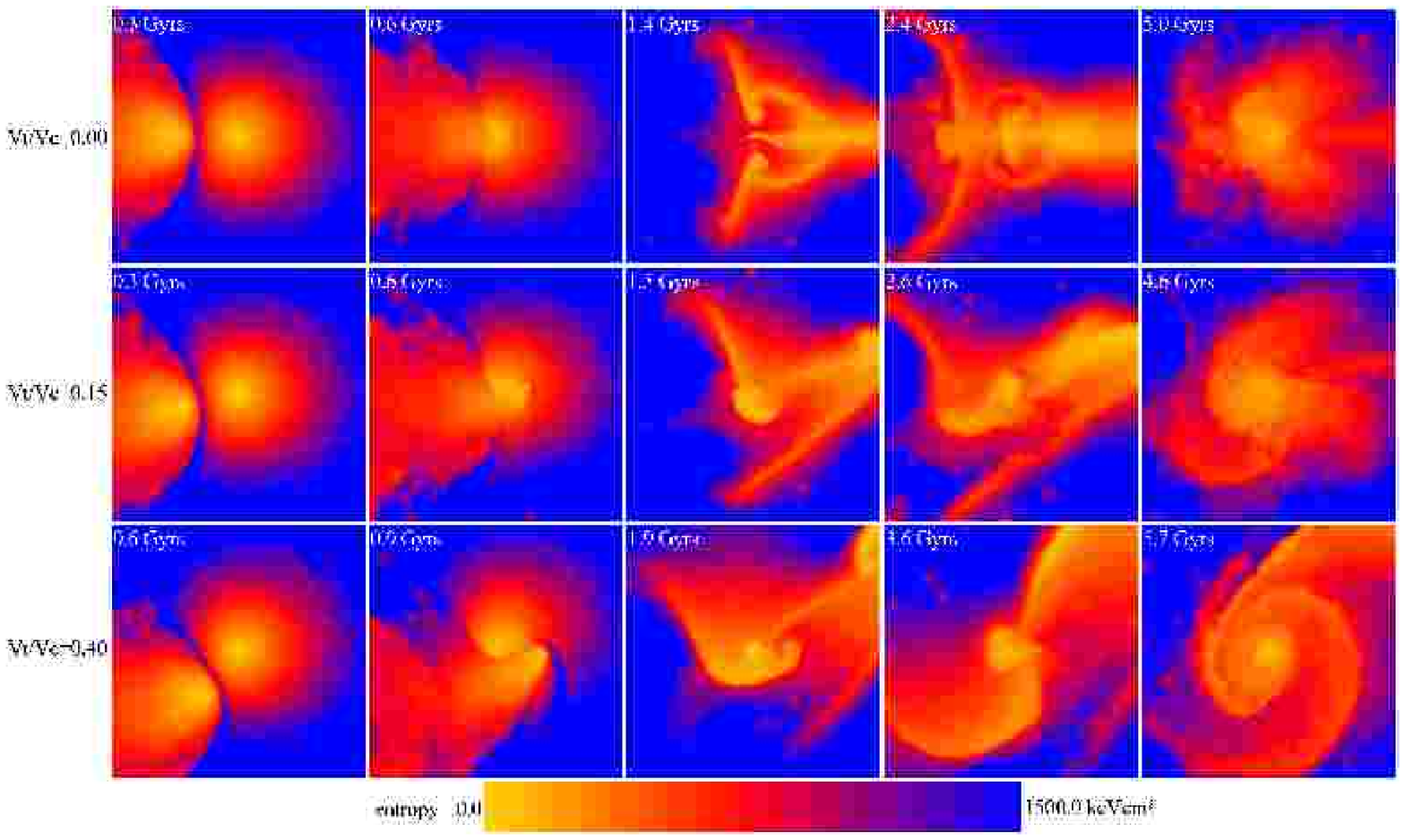}
\caption{Maps for a $0.5$\Mpc\ thick slice ($3$\Mpc\ on a side) through the centres of our $3:1$ simulations$^\dagger$.  The times represented are the same as Fig. \ref{fig-3to1_Sigmagas}.}
\label{fig-3to1_entropy}
\end{center}
\end{minipage}
\end{figure*}

Throughout our analysis we shall refer to three coordinate axes; $x$ and $y$ will denote directions in the plane of the initial orbit and $z$ the direction orthogonal to this plane.  To properly capture the initial effects of the encounter as the outermost envelopes of the systems make contact and the interaction shocks begin to form, we initialise our clusters separated along the $x$ axis so that their smoothly truncated envelopes barely touch (\ie separated by slightly more than the sum of their virial radii).  As a result, the initial conditions we have outlined above must be imposed at a time part way through the run when the centre of mass of the secondary system crosses the virial radius of the primary.  To determine the initial velocities which yield these conditions, we first approximate the systems as point masses and analytically determine the initial conditions which yield the desired constraints.  To account for the extended, tidally deforming mass distributions, we then iteratively refine these orbits with low-resolution dark-matter-only simulations.

\begin{figure*}
\begin{minipage}{175mm}
\begin{center}
\epsfysize=95mm \epsfbox{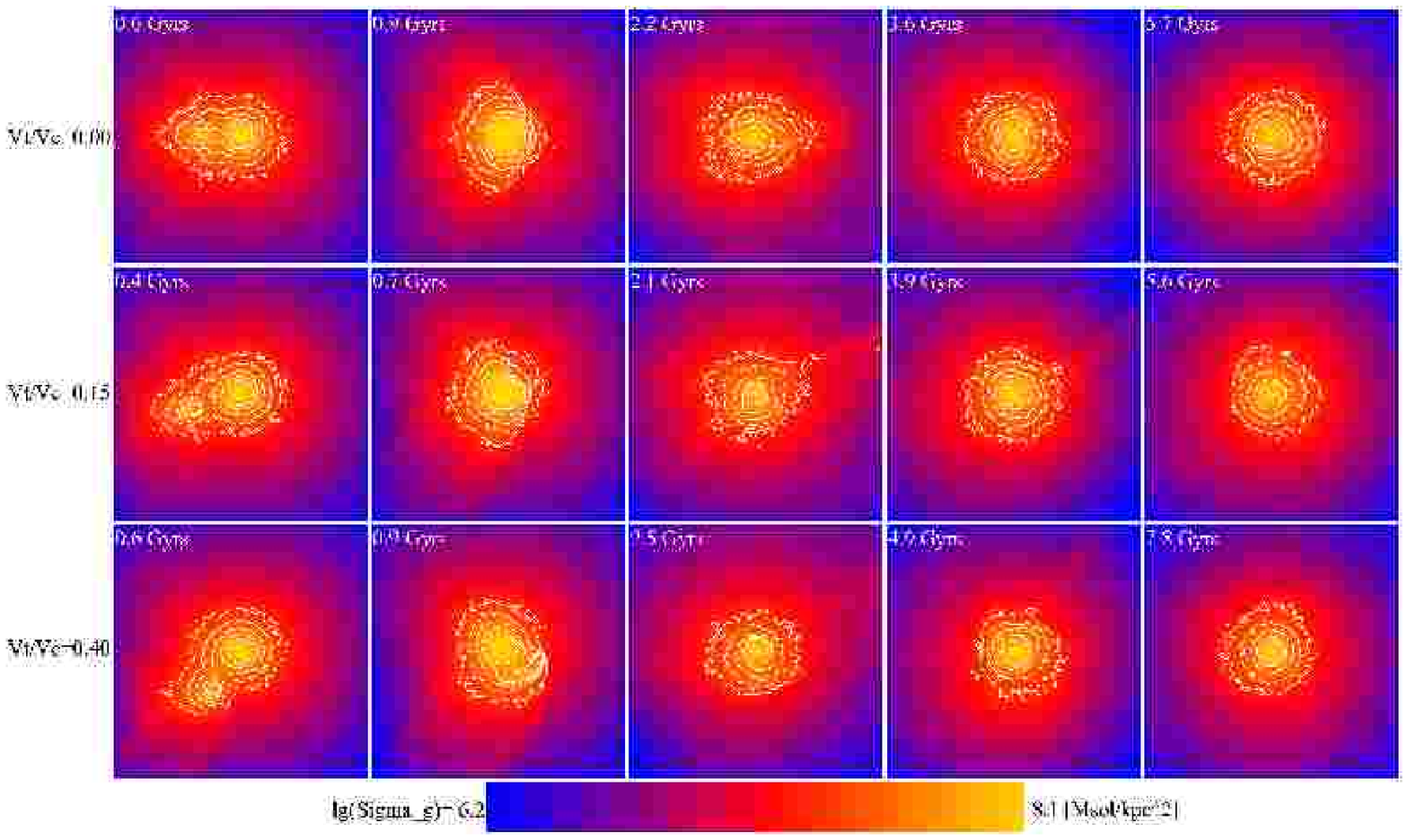}
\caption{Maps of the gas surface density for our 10:1 simulations, projected along an axis normal to the plane of the orbit ($3$\Mpc\ on a side).  White contours are X-ray surface brightness isophotes from simulated 50\ks\ $z=0.1$ \Chandra\ observations ($3\times$ to $48\times$ the background stepping by factors of $2$).  Times depicted are \tclosest$-200$\Myrs, \tclosest$+100$\Myrs, \tapo, \taccrete$+100$\Myrs~and \trelax. See Figure 1 caption or Table 2 for a brief definition of the timescales. All times are measured relative to the $t_o$ when the secondary's centre of mass crosses the  $R_{200}$ radius of the primary for the first time$^\dagger$.}
\label{fig-10to1_Sigmagas}
\end{center}
\end{minipage}
\end{figure*}

Table \ref{table-initial_conditions} lists the parameters relevant to the initial conditions of our runs including the centre-of-mass separation of the secondary from the primary ($r$), initial tangential and radial centre-of-mass velocities of the secondary relative to the primary ($v_t$ and $v_r$), the time of virial crossing ($t_o$, using the virial radius $R_{200}$), and the centre-of-mass distances of closest approach ($r_{min}$).  As a test of the validity of our orbits we have compiled the spin parameter and dark matter triaxial shapes for our final combined systems.  We compute the spin parameter from $\lambda = J |E|^{1/2} / G M^{5/2}$ with $J$ being the total angular momentum, $E$ the total energy and $M$ the total mass of the remnant, all integrated within $R_{200}$.  We compute the triaxial axes ratios at $R_{200}$ (denoted $q_{200}$ and $s_{200}$) following the approach of \citet{DubinskiandCarlberg91}.  All are in reasonable agreement with accepted distributions from cosmological simulations \citep{BarnesEfstathiou87,JingSuto02}.

\subsection{Numerical methods}

To compute the hydrodynamic forces between gas particles, we use GASOLINE in its default configuration \citep{wadsley04}:  we use the Benz arithmetic-asymmetric implementation of the SPH momentum and energy equations.  The gas quantities are smoothed over 32 particles,  the Courant time step parameter is set to $\eta_c = 0.4$, and the viscosity uses the standard Monaghan formulation of viscosity with $\alpha = 1$ and $\beta = 2$.

\begin{figure*}
\begin{minipage}{175mm}
\begin{center}
\epsfysize=95mm \epsfbox{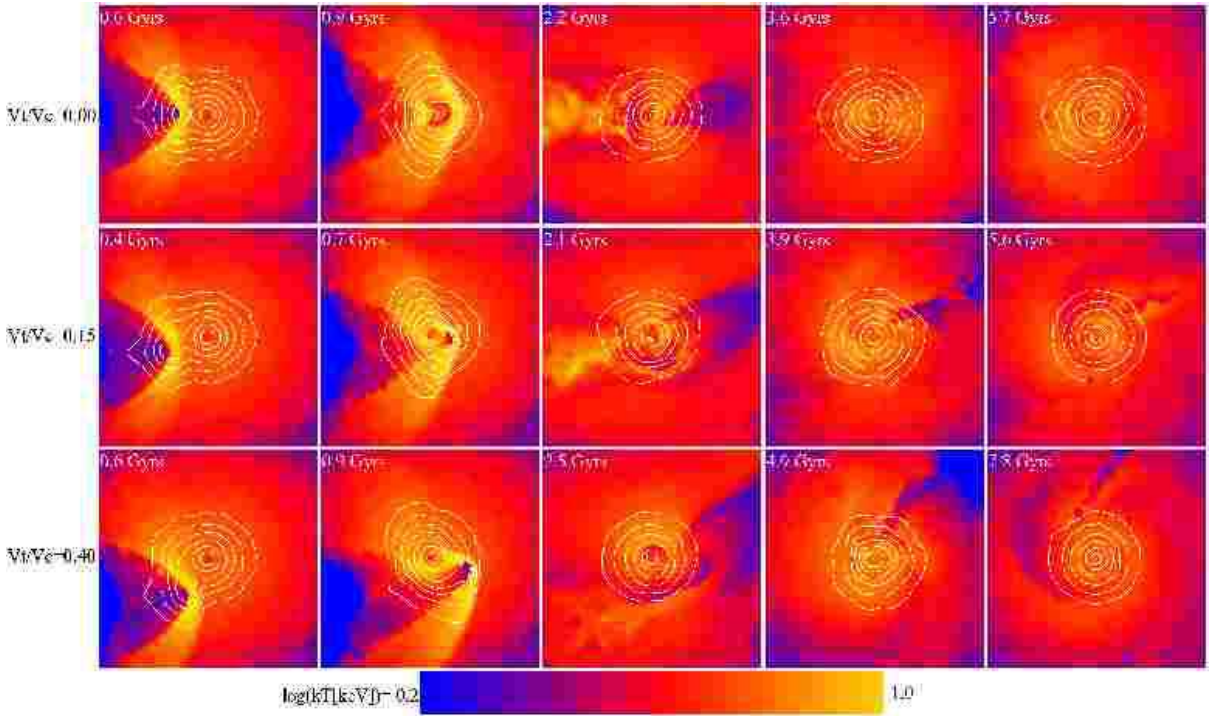}
\caption{Maps of the projected ``spectroscopic-like'' temperature \citep{Mazzottaetal04} maps for a $0.5$\Mpc\ thick slice ($3$\Mpc\ on a side) passing through the centre of our $10:1$ simulations.  Contours depict the SZ effect (for the entire simulation volume along the line of sight) and represent $log(y)=-5.5$ to $-4$ in increments of $0.25$$^\dagger$.  The times represented are the same as Fig. \ref{fig-10to1_Sigmagas}.}
\label{fig-10to1_Tsl}
\end{center}
\end{minipage}
\end{figure*}
\begin{figure*}
\begin{minipage}{175mm}
\begin{center}
\epsfysize=95mm \epsfbox{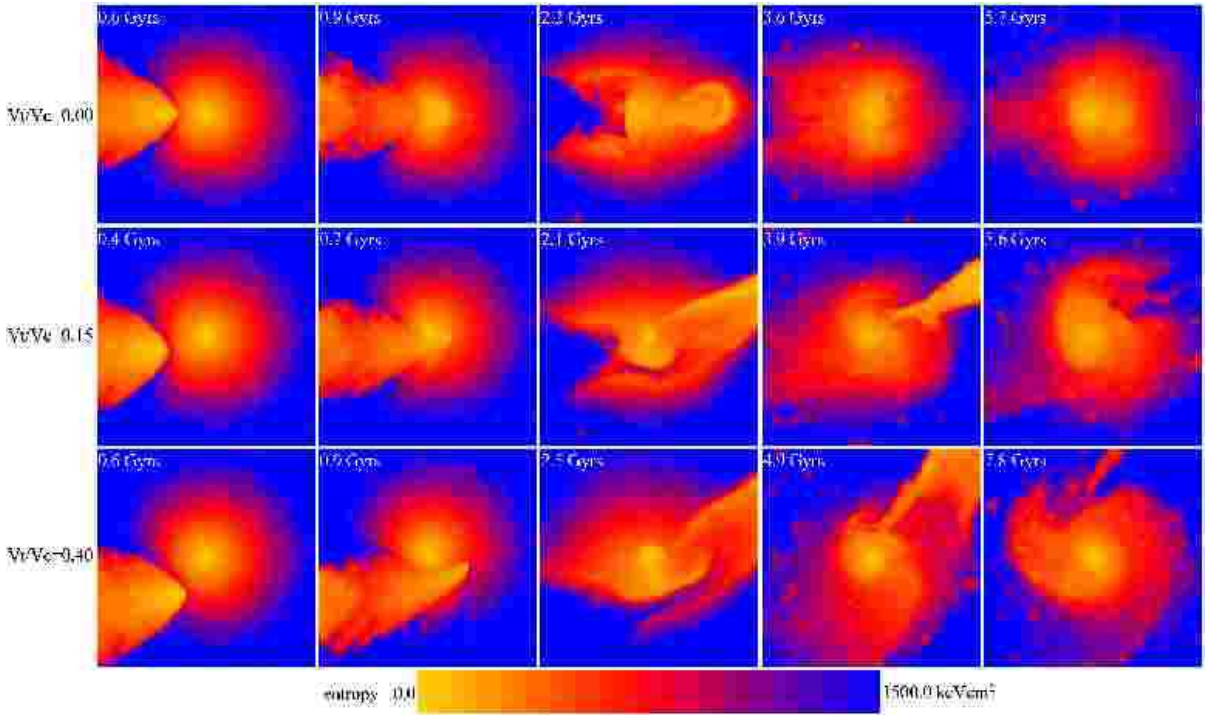}
\caption{Maps for a $0.5 \Mpc$ thick slice ($3 \Mpc$ on a side) through the centres of our $10:1$ simulations$^\dagger$.  The times represented are the same as Fig. \ref{fig-10to1_Sigmagas}.}
\label{fig-10to1_entropy}
\end{center}
\end{minipage}
\end{figure*}

The gravitational forces are calculated with a tree that uses an opening angle $\theta=0.8$.  Each particle trajectory is integrated on its own time step $\Delta t_i$, which is related to the particle's acceleration $a_i$ by the criterion $\Delta t_i < \eta (\epsilon / a_i)^{1/2}$ with tolerance parameter $\eta = 0.2$.  The gravitational softening uses spline kernel interpolation, and the spline softening length is set to a constant value of $\epsilon = 10\kpc$\ for all cluster particles.  This ensures that gravitational forces in the core are well resolved.

\begin{table*}
\begin{minipage}{170mm}
\caption{Times of important stages in the dynamical evolution of our simulations measured relative to $t_o$, the time when the secondary's centre of mass crosses the primary's $R_{200}$ radius.  These are as follows:  $t_{closest}$ is the time of first pericentric passage; $t_{apo}$ is the time of the first apocentric passage; $t_{accrete}$ is the time of the second pericentric passage typically during which the secondary's gas core merges with that of the primary and it also marks the beginning of stream accretion; $t_{relax}$ is the time at which the total system visually appears relaxed in simulated $50$\ks\ \Chandra\ observations at $z=0.1$; $t_{virial}$ is time at which the system's dark matter is assessed to be virialised within  $R_{500}$; and $t_{hydro}$ is the time at which the system returns to hydrostatic equilibrium (to within $10$\%) at $R_{500}$.    
Quantities in brackets indicate the redshift at which the secondary core would have to cross $R_{200}$ of the primary for the system to be observed in the given state at 
$z=0.1$ and $z=0.5$ respectively.  Additional parameters in the table are: the mass ratio of the primary to the secondary (column 1) and the secondary's transverse velocity at
when its centre of mass crosses the virial radius of the primary, in units of the primary's circular velocity (column 2). 
\label{table-times}}
\begin{tabular}{cccccccc}
\hline
$M_p$:$M_s$      &
$v_t/v_c$        & 
$t_{closest}$    & 
$t_{apo}$        & 
$t_{accrete}$    & 
$t_{relax}$      &
$t_{virial}$     &
$t_{hydro}$      \\
\hline
1:1   & 0.00 & 0.5 (0.14, 0.58) & 1.2 (0.21, 0.70) & 2.2 (0.33, 0.92) & 4.4 (0.65, 1.73) & 4.4 (0.65,1.73) & 7.9 (1.73,$>$10)\\ 
1:1   & 0.15 & 0.6 (0.15, 0.60) & 1.5 (0.25, 0.77) & 2.3 (0.34, 0.95) & 4.4 (0.65, 1.73) & 4.4 (0.65,1.73) &       --        \\ 
1:1   & 0.40 & 0.5 (0.14, 0.58) & 1.4 (0.24, 0.75) & 2.7 (0.39, 1.06) & 5.4 (0.85, 2.44) & 3.8 (0.55,1.44) & 3.3 (0.47,1.25) \\ 
3:1   & 0.00 & 0.5 (0.14, 0.58) & 1.4 (0.24, 0.75) & 2.3 (0.34, 0.95) & 5.0 (0.77, 2.10) & 3.4 (0.49,1.28) & 4.3 (0.63,1.67) \\ 
3:1   & 0.15 & 0.5 (0.14, 0.58) & 1.5 (0.25, 0.77) & 2.5 (0.26, 1.00) & 4.6 (0.69, 1.84) & 3.4 (0.49,1.28) & 4.3 (0.63,1.67) \\ 
3:1   & 0.40 & 0.8 (0.18, 0.63) & 1.9 (0.29, 0.85) & 3.5 (0.50, 1.32) & 5.7 (0.92, 2.75) & 4.3 (0.63,1.67) & 3.6 (0.52,1.36) \\ 
10:1  & 0.00 & 0.8 (0.18, 0.63) & 2.2 (0.33, 0.92) & 3.5 (0.50, 1.32) & 4.8 (0.73, 1.96) & 4.6 (0.69,1.84) & 5.2 (0.81,2.26) \\ 
10:1  & 0.15 & 0.6 (0.15, 0.60) & 2.1 (0.32, 0.90) & 3.8 (0.55, 1.44) & 5.6 (0.90, 2.65) & 4.5 (0.67,1.78) & 5.1 (0.79,2.18) \\ 
10:1  & 0.40 & 0.8 (0.18, 0.63) & 2.5 (0.26, 1.00) & 4.8 (0.73, 1.96) & 7.8 (1.68,$>$10) & 5.6 (0.90,2.65) & 5.8 (0.95,2.88) \\
\hline
\end{tabular}
\end{minipage}
\end{table*}

Radiative cooling allows for the formation of resilient dense gaseous cores.  In some cases it can also dissipate significant amounts of the thermal energy deposited in the centres of merger remnants, further stabilising them.  Since we are acutely interested in the evolution of these regions for this series of papers, the inclusion of cooling in our simulations is necessary.

We focus our analysis on radii $r>40 \kpc \sim 0.03 R_{500}$ and have sought to ensure that we accurately model the behaviour of the system in that range.  This is because the precise nature of very central regions of the observed compact cool cores is still not well understood.  The temperature floors observed in their centres \citep{Peterson03} indicate the presence of a source of heating whose nature is still highly uncertain with several candidate mechanisms presently being considered (\eg heating from AGN, conduction and turbulent mixing to name only a few).  There is also evidence of other complex phenomena occurring in the central $40$\kpc, such as multiphase cooling and ongoing star formation \citep{Jaffeetal05,Egamietal06}.  These processes cannot be realistically captured with SPH simulations such as ours.  In light of these issues, we shall compute ``central'' quantities at $40 \kpc$ and shall exclude the central $30 \kpc \sim 0.025R_{500}$ of the system from integrated quantities such as luminosities and globally averaged temperatures.  

For our simulations we use a cooling function appropriate to a primordial gas.  However, typical clusters have metallicities of $Z\sim0.3Z_\odot$ and by omitting heavy element line emission, we are underestimating the cooling rate of material with temperatures $T = 10^5$--$10^7\K$.  A very small fraction ($<1\%$) of the gas mass in our systems exists at these temperatures; most of the mass has $T=10^7$--$10^8$\K.  Additionally, this discrepancy has a negligible effect on the dynamics of our simulations.   All subsequent analyses (\eg producing X-ray surface brightness maps) are computed assuming a more realistic Raymond and Smith model \citep{RaymondSmith77} compiled for a metallicity of $Z\sim0.3Z_\odot$.

We include star formation in our simulations using an algorithm which turns cold and dense gas into collisionless star particles, using a recipe similar to that of \citet{KWH96} and \citet{stinson06}.  We first select gas particles which are capable of forming stars on the basis of four eligibility criteria: the gas must be dense ($n_H > 0.1 \cm^{-3}$); it must be either cool ($T < 3 \times 10^4 \K$) or cooling ($dT/dt < 0$); its flow must be converging ($\nabla \cdot \vec{v} < 0$); and the particle must be Jeans unstable.  Once selected, star formation produces star particles with a probability $p = 1.0 - \exp(-c_\ast \Delta \! T /t_{\it form})$ where the star-formation efficiency is taken to be $c_\ast = 0.1$.  We set $t_{\it form}$ to the maximum of the cooling time and the dynamical time, unless the particle is already cool ($T < 3 \times 10^4 \K$) in which case we always set it to the dynamical time.  If the particle is chosen to form stars, a new star particle is created with $1/3$ of the mass of a full-sized gas particle.  

Star formation and supernova explosions go hand in hand.  The energy released by the supernovae is injected into the surrounding intracluster medium as follows.
At every timestep, we consider each star particle in the simulation and based on its age, determine the associated supernovae rate.   The total energy released by the 
supernovae over the course of the timestep is then injected at a steady rate to the surrounding gas particles over the timestep.  The star formation and feedback recipe we use is similar to that described in \citet{stinson06}, except that we do not disable radiative cooling during feedback, and stellar winds which release mass from intermediate-mass stars are not included.
This is a ``minimal'' feedback scheme, since the energy is quickly radiated away and the feedback does not impact the evolution of the ICM in any meaningful fashion.

Finally, we note that our chosen mass resolution models the initial conditions of a $10^{15} M_\odot$ system with 217441 dark matter and 211914 gas particles, where about 70\% of the particles in each species are within $R_{200}$. The dark matter particle mass is $4.4\times10^9 M_\odot$ and the initial gas particle mass is $5.9\times10^8 M_\odot$.  In test runs, we have found that the cooling rates of isolated clusters are not robust until there are $\tsim 100$ particles within the ``cooling radius'' (\ie the radius within which $t_{cool}<t_{\it Hubble}$; $R_c\sim150$\kpc\ for our simulations).  With our initial conditions, the central core ($r_c\sim50$\kpc) of a $10^{15} M_\odot$ system is initially resolved by $\sim 240$ particles, and its cooling radius by $\sim 6300$ particles. 

\subsection{Simulated observations}\label{numerics-observations}

In Section \ref{analysis-disturbed} we will examine possible discrepancies between the relaxed appearance of systems and the degree to which they are formally so.  For this reason as well as to determine the observability of features in the projected mass, temperature and entropy maps presented in Figs. \ref{fig-1to1_Sigmagas} to \ref{fig-10to1_entropy}, we have created mock \Chandra\ observations of our simulations.  To create these we generate 325 projected X-ray maps ranging from $0.5$\keV\ to $7.0$\keV\ in $20$\eV\ intervals using the Theoretical Image Processing System (TIPSY).  This package takes the SPH outputs from our GASOLINE simulations and produces smoothed projected X-ray surface brightness maps with the appropriate variable SPH kernel applied individually to the flux represented by each particle.  To these we add a three component X-ray background (following the prescription of the Quicksim \XMM\ simulation package) consisting of an extragalactic power-law component \citep{Chenetal97} and two thermal components of $10^6 \K$ and $10^{6.6} \K$ modelled with a Raymond and Smith plasma \citep{RaymondSmith77}.  Generally, each image would then be multiplied by a energy-dependent radially varying effective area.  However, we are interested in the detectability of extended features at $z=0.1$ and our merging systems rarely fit within the \Chandra\ field of view (FOV) at such a redshift.  For this reason, we simulate a ``mosaic'' mode for our simulated observations using constant but energy dependent effective areas, area averaged over the \Chandra\ FOV.  An unvignetted and spectrally flat particle background is then added with normalisation $3.5\times10^{-2}$ cts~arcmin$^{-2}$s$^{-1}$\keV$^{-1}$.  The resulting images are convolved with an azimuthally symmetric energy dependent PSF taken from the on-axis PSFs specified by version 2.23 of the \Chandra\ CALDB.    The result is converted to a photon flux at a redshift of $z=0.1$, quantised with a Poisson distribution, and co-added.  For this paper, we use an integration time of $50$\ks.


\section{Qualitative evolution}\label{sec-analysis}

In what follows we present a qualitative account of common elements in the dynamical evolution of typical cluster mergers.  A series of stages which mergers generically progress through will be identified and the evolution of the system's physical and observable properties described.  In the process, several classes of transient structures will be introduced.  This discussion is kept brief and is intended to provide a context for the discussion of the observability of substructure and relaxation which follows in Sections \ref{analysis-disturbed} and \ref{analysis-relaxation}$^\ddagger$\renewcommand{\thefootnote}{\fnsymbol{footnote}}\footnotetext[3]{Readers are invited to peruse relevant digital movies detailing the evolution of various system properties over the course of the simulations at \URL}\renewcommand{\thefootnote}{\arabic{footnote}}.  For those seeking a more detailed description, a specific discussion of each class of transient structure can be found afterwards in Section \ref{analysis-transients}.  All quoted times are measured from $t_o$, the moment at which the secondary system traverses $R_{200}$ of the primary.

\subsection{Evolutionary stages}\label{analysis-generic}

All of the simulations we have studied proceed through a similar evolutionary sequence involving five distinct stages: a pre-interaction phase, first core-core interaction, apocentric passage, secondary core accretion, and relaxation.  Several important times during this progression which we shall refer to throughout our analysis are listed in Table \ref{table-times}. 

Initially, our systems are constructed to possess small core radii ($r_c \sim 50$\kpc, measured from $\beta$-model fits), low central entropies ($S=10$\keVcmsq\ at $10$\kpc) and strong central positive temperature gradients.  The low-density outer atmosphere of each system becomes distorted during pre-interaction through tidal forces and compression, adiabatically raising the temperature of material between and producing a short-lived bridge in surface brightness joining them.  As the cores continue to accelerate upon approach, a pair of shock fronts materialise and are driven towards each core, heating and compressing them briefly.  The effects of this interaction reaches its maximum strength at \tclosest$=0.5-0.8$\Gyr\ when the cores reach their closest approach.  The system's temperature and luminosity increase sharply for $\sim 400$\Myrs\ at this point.  Cooling of gas to a cold ($T<2.5\times10^4$\K) state or to stars is subsequently quenched for $2-3$\Gyrs\ in most cases but is merely suppressed in the off-axis 10:1 interactions.  Meanwhile, the primary system's projected central positive temperature gradient is strongly reduced and its core radius increased.  A detailed analysis of these phenomena is presented in the second and third papers of this series \citep[][in preparation]{P06b,P06c} in which the effects of mergers on the global properties of merging systems and the structure and appearance of compact cool cores will be studied.

In every case (including the head-on collisions) some part of the secondary's cool core survives its first encounter with the primary core, forming a single clump and large cool stream of stripped material in off-axis cases and strings of one or more clumps in head-on cases.

At $t_{apo}=1.2-2.5$\Gyrs\ the disturbed secondary core reaches a maximum separation of $1.2$-$1.7$\Mpc\ from the primary core.  The surviving portion then returns for another encounter with the primary, reaching a second pericenter at \taccrete$=2.2-4.8$\Gyrs.  In Table \ref{table-times} we see that the time elapsed until either of these stages generally correlates with the mass ratio and impact parameter of the interaction, presumably due to the declining efficiency of dynamical friction.  Second pericentric passage marks the beginning of an extended period of several \Gyrs\ during which material dispersed from the secondary core accumulates onto the core of the primary system.  Although a small portion of the secondary core can survive the resulting disruption and experience tertiary encounters with the primary, there are no realistically observable traces of it following \taccrete\ in any of the cases we have studied$^7$\footnotetext[7]{Higher impact parameters than probed by our simulations, although atypical, could result in longer lived distinct secondary cores.}.  For this reason, our systems are considered to be evolving as a single merger remnant from \taccrete\ onwards.

In a study of the orbital parameters of cluster substructure in a cosmological context (including gas but not cooling), \citet{Tormenetal04} similarly find that the gaseous component of secondary systems becomes disrupted shortly after second pericentric passage.  They also present simple analytic models for predicting the time and distance of first pericentric and apocentric passages.  Our simulations compare well to their model predictions, supporting the consistency of our initial conditions with the orbital properties of substructure in cosmological simulations.

Following the accretion of the secondary core, the resulting remnant proceeds through a period of relaxation.  It appears undisturbed under reasonable observational circumstances in simulated \Chandra\ observations by \trelax$=4.4-7.8$\Gyrs\ but past this time, continues to evolve until the end of our simulations:  recovery of the core's central temperature decrement is still proceeding \citep[see][for more details]{P06c} as well as accretion of both dark matter and gas dispersed beyond the $R_{200}$ radius during the merger.  The system sustains a virialised state within $R_{500}$ after $t_{virial}$ and hydrostatic equilibrium at $R_{500}$ after $t_{hydro}$ (see Section \ref{analysis-relaxation}).

Following \tclosest, short-lived structures in the system's distribution of X-ray surface brightness, SZ signature, temperature and entropy arise.  In Figs. \ref{fig-1to1_Sigmagas} to \ref{fig-10to1_entropy} we present projected gas surface density/X-ray brightness, temperature/SZ and entropy maps of each of our simulations near several of the times discussed above.  In what follows we shall use these figures to present the progression of these structures (both physical and observable) as they arise and dissolve.  

{\bf Readers may find it useful to refer to movies of our simulations$^\ddagger$ while reading the following subsections.}

\subsection{Evolution of head-on mergers}\label{analysis-head_on}

Our head-on mergers evolve in a distinctly different way from our off-axis mergers, with even their qualitative evolutions exhibiting strong dependences on the mass ratio of the systems.  We thus isolate the account of their evolution from that of the off-axis cases.

\subsubsection{Physical state}

In our equal mass head-on merger, two near-planar shocks are produced shortly before \tclosest\ and are driven into and through each core (in good agreement with RS01).  They subsequently exit the system as two symmetric hemispherical fronts.  Due to the confining geometry of this situation, material initially in the cores of the merging systems remains in relatively high density structures during the impact.  From \tclosest\ to approximately \tapo, much of this material occupies an expanding disk oriented normal to the axis of the merging systems' motion and a filament of dense clumps strung between the remnants of the two cores, both of which are clearly visible in Fig. \ref{fig-1to1_Sigmagas}.  In Figs. \ref{fig-1to1_Tsl} and \ref{fig-1to1_entropy} we see that this material is cold ($T\sim 1.0$\keV) and has low entropy ($S\sim 10$\keVcmsq).  At approximately \taccrete\ it reaccretes to the remnant core having radiated most of the thermal energy generated by the impact.  As a result, the system's core returns to a state of cooling twice as efficient at converting the hot ICM into stars and condensed gas ($T<2.5\times10^5$K) as the primary prior to the merger.  Although equal mass head-on mergers are predicted to be rare in CDM models of structure formation$^{8}$\footnotetext[8]{However, they are not without observational precedence; RXCJ0532.9-3701 \citep{Finoguenovetal05} has boxy isophotes and a cloverleaf morphology in entropy, much like our relaxing equal mass head-on merger}, this supports the findings of \citet{Motl04} who find that mergers can act as a means of constructing actively cooling cores.  We will study this in more detail in \citet{P06c}.

Comparing Fig. \ref{fig-1to1_Sigmagas} to Fig. \ref{fig-3to1_Sigmagas} we can see that reducing the mass of the secondary to a third of the primary's dramatically changes even the qualitative behaviour of a head-on collision.  Although two shocks are still produced, the one leading the secondary's motion is much more significant to the system's evolution.  As Figs. \ref{fig-3to1_Tsl} and \ref{fig-3to1_entropy} show, the bow-shaped geometry of this shock (again, in good agreement with RS01) displaces the material of the primary core forward and laterally.  The result is a conical shell of cool ($T\sim 3$\keV) moderate-entropy ($S\sim 150$\keVcmsq) gas, entrained by the remainder of the secondary core at its apex.  This structure surrounds a bubble of hot ($T\sim10$\keV) high-entropy ($S\sim550$\keVcmsq) gas.  These figures also illustrate how the low pressure region created in the wake of the secondary's motion is quickly and convectively filled by this displaced material.  Material displaced most laterally reaccretes first, flowing in through the back of the secondary's wake.  Material initially displaced more forwardly is dragged in behind afterwards.  Upon returning to the core, this material experiences additional shocking, resulting for example in the high temperature ($T\sim 9.5$\keV) region seen to the left of the core shortly after \taccrete\ in Fig. \ref{fig-3to1_Tsl}.

In the 10:1 case, the secondary core penetrates the primary core and deposits nearly all the kinetic energy of its gaseous component into the core of the primary.  The result is a large ($\sim 500$\kpc) bubble with an expanding shell of cool ($T\sim 3.5$\keV) low-entropy ($S\sim 70$\keVcmsq) material filled with hot ($T\sim 7$\keV) moderate-entropy ($S\sim 180$\keVcmsq) gas.  In Fig. \ref{fig-10to1_Tsl} the initial formation of the bubble after \tclosest\ is illustrated.  This bubble lasts for $\sim 1.5$\Gyrs\ and is carried forward to a radius of $\sim 750$\kpc\ by the momentum of the collision. Its final extent can be seen in Figs. \ref{fig-10to1_Tsl} and \ref{fig-10to1_entropy}.  Shortly before \tapo\ it stops moving and expanding, collapses into a low entropy stream and accretes back to the core by \taccrete.

\subsubsection{Observable state}

Following the procedure detailed in Section \ref{numerics-observations} we have produced simulated \Chandra\ observations and overlaid their contours onto the projected gas surface density maps presented in Figs. \ref{fig-1to1_Sigmagas}, \ref{fig-3to1_Sigmagas} and \ref{fig-10to1_Sigmagas}.  This allows us to determine which distributed features have sufficiently high surface brightnesses to be detected under reasonable observational circumstances.  We have also generated SZ maps and overlaid their contours on the temperature maps presented in Figs. \ref{fig-1to1_Tsl}, \ref{fig-3to1_Tsl} and \ref{fig-10to1_Tsl}.  Since these maps do not include the effects of instrumentation, we cannot comment in detail on the detectability of low surface brightness features through SZ observations.  We place our faintest contours at a level of $\log{y}=-5.5$ \citep[$\delta T=-17\mu K$ at 30GHz,][]{McCarthyetal03} to approximate the depth to which the next generation of SZ imaging instruments will reach \citep[\eg $-10\mu K$ for SZA at 30GHz,][]{Knoxetal04}.  The resulting SZ limit is a good match to the system extent realistically observable by \Chandra.

In the 1:1 case, we can see in Figs. \ref{fig-1to1_Sigmagas} and \ref{fig-1to1_Tsl} that both the expanding disk and filament of disturbed core material created in the impact and discussed above are reasonably observable through X-ray observations.  Comparing these figures we can see that the remnant's X-ray isophotes are extremely boxy while the SZ contours are significantly more regular and much less elliptical (this is a trend that we observe in all of our simulations).  Complicated temperature and entropy variations are seen throughout the duration of the merger, with small $\sim100$\kpc\ scale variations ($\Delta T/T\sim 15\%$) remaining at \trelax.

In the 3:1 head-on case, Figs. \ref{fig-3to1_Sigmagas} and \ref{fig-3to1_Tsl} illustrate that the extended features generated in this merger following \tclosest\ are similarly observable as in the 1:1 case.  However, we can see that at \taccrete\ there are large extended low surface brightness features that escape detection by \Chandra.  Complicated temperature and entropy variations, distinctly different from those observed in the 1:1 case, emerge after \taccrete\ and persist until the end of the simulation.  The temperature fluctuations present at \trelax\ are more significant ($\Delta T/T\sim 20\%$) and spatially irregular than in the 1:1 case. 

Lastly, Figs. \ref{fig-10to1_Sigmagas} and \ref{fig-10to1_Tsl} illustrate the observability of extended structures formed in our 10:1 head-on merger.  We can see from these that the large expanding bubble formed from the collision is not detectable when it reaches its largest extent at \tapo.  Furthermore, the system looks quite regular in both X-ray and SZ at \taccrete, although small but significant irregularities in the central X-ray isophotes persist for $\sim 2$\Gyrs\ afterwards.  Temperature and entropy fluctuations are present throughout the merger with the bubble formed from the core material dominating the temperature and entropy distributions after \tclosest.  Significant fluctuations in temperature ($\Delta T/T\sim 20\%$) are present at \trelax, with little change occurring between \taccrete\ and \trelax.

\subsection{Evolution of off-axis mergers}\label{analysis-off-axis}

In all of the off-axis simulations we have studied, several distinct classes of transient structure are formed when the secondary (and sometimes primary) system's core becomes disturbed.  Interestingly, the qualitative evolution of these structures is remarkably independent of impact parameter and mass ratio over the ranges we have studied.  Furthermore, their properties are consistent with those of ``cold fronts'' observed in many \Chandra\ and \XMM\ observations \citep{Markevitchetal00}.  Cold fronts are contact discontinuities between regions of bright, dense, cold gas and faint, hot, rarefied gas.  These regions are observed to be in near pressure equilibrium with the surrounding ICM and the jumps in temperature across their interface imply mildly sonic or sub-sonic velocities.

Several authors have studied simulations of merging clusters illustrating that the ram pressure disruption of cold low entropy cores during mergers can account for these features \citep{Bialeketal02,Heinzetal03,Mathisetal05}.  Once removed from the dark matter potential confining it, stripped gas lags behind its dark matter and galactic components, adiabatically expanding and cooling until it establishes pressure equilibrium within the resulting merger remnant.

Our simulations confirm these results.  In what follows we will give a qualitative account of our off-axis mergers' evolution through a series of transient structures and discuss their observable properties.  We shall find that there are many points when several classes of structure resembling cold fronts can be created.  Additional details can be found in Section \ref{analysis-transients} where each of these classes will be studied in detail and new mechanisms involved in the production and evolution of cold fronts will be presented.

\subsubsection{Physical state}

In all of our 1:1 and 3:1 off-axis mergers, ram pressure stripping causes both the primary and secondary cores to develop ``comet-like'' head/tail morphologies in gas surface density, X-ray surface brightness, temperature and entropy during their first pericentric passage.  In the 10:1 cases, this is true only for the secondary core.  In all cases, this morphology starts $\sim 100$\Myrs\ before \tclosest\ and ends shortly before \tapo\ with a typical duration of $\sim0.5$\Gyrs.

Upon reaching apocentric passage, several extended structures in the system's distribution of temperature, entropy and gas density have evolved.  In Figs. \ref{fig-1to1_Sigmagas} and \ref{fig-3to1_Sigmagas} we see that a bridge connecting the surviving portions of the primary and secondary cores has formed in the 1:1 and 3:1 cases (see Section \ref{analysis-bridges}).  Several long and roughly radial outwardly moving plumes of cool ($T\sim3$\keV) low-entropy ($S\sim60$\keVcmsq) material have also formed (see Section \ref{analysis-plumes}).  These features are not obvious in the system's gas density or surface brightness maps, but are evident in the temperature and entropy maps presented in Figs. \ref{fig-1to1_Tsl}, \ref{fig-1to1_entropy}, \ref{fig-3to1_Tsl} and \ref{fig-3to1_entropy}.  All of these structures are essentially absent in our off-axis 10:1 mergers.

\begin{figure*}
\begin{minipage}{175mm}
\begin{center}
\leavevmode \epsfysize=10cm \epsfbox{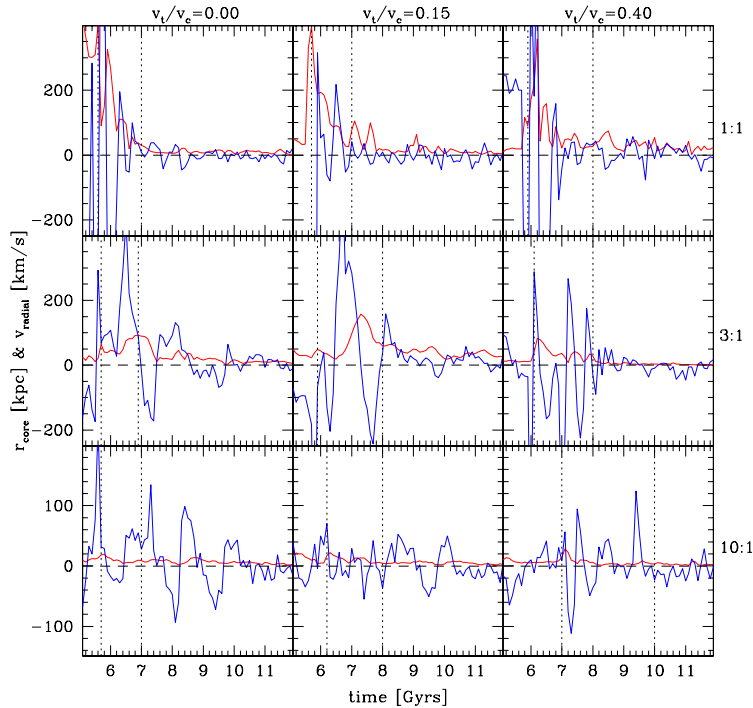}
\caption{Displacement of the centre of mass of gas within the central $150$\kpc~from its dark matter centre of mass (smoother, lower-amplitude red curve) and the mass-averaged velocity of the same material in the direction of that displacement (the more erratic blue curve) for the late stages of each of our simulations.  Vertical dashed lines correspond to \taccrete\ and \trelax.  The increased activity between these two timescales is primarily the result of the gas core being disturbed by the secondary's gas core during the second and typical the final pericentric passage, and by the ensuing cold stream.  Values across the top indicate the transverse velocity at $t_o$ depicted by each column while text on the right indicates the mass ratio depicted by each row$^\dagger$.}
\label{fig-rv_t}
\end{center}
\end{minipage}
\end{figure*}

Following \tapo, material displaced as plumes from the primary and secondary cores stops expanding away from the system and begins returning towards the remnant core.  At \taccrete, the secondary core returns for its second (and last) pericentric passage of the system.  After this, plume material begins accreting to the remnant core as collimated high velocity ($\sim 1000$\kms) streams of material (see Section \ref{analysis-streams}).  At this point, evidence of a secondary core has been destroyed and these features are not associated with an apparent merging system.  Stream material is generally shock heated as it accretes to the remnant core.  These streams occur even in our 10:1 cases and are visible in the temperature and entropy maps presented in Figs. \ref{fig-1to1_Tsl}, \ref{fig-1to1_entropy}, \ref{fig-3to1_Tsl}, \ref{fig-3to1_entropy}, \ref{fig-10to1_Tsl} and \ref{fig-10to1_entropy} as narrow and radially oriented cold low entropy structures following \taccrete.  When they accrete directly to the remnant core, these streams lead to multiple discontinuities, or ``edges'' in the system's central surface brightness distribution (see Section \ref{analysis-edges}).

\subsubsection{Observable state}

Elongated and curved disturbed cores have been observed in several systems (\eg A85, A1758, and A2256) and are frequently attributed to the disruption of accreting systems due to ram pressure stripping.  We observe this for only a short duration ($\lesssim 0.5$\Gyr) following \tclosest\ in all of our off-axis mergers.  Prior to \tclosest, this morphology does not manifest.  At \tapo, longer lived elongated tails are created but the ejection mechanism is not ram pressure stripping (see Section \ref{analysis-plumes}).

In Fig. \ref{fig-1to1_Sigmagas} we see that the bridge formed after \tclosest\ in our off-axis 1:1 mergers remains visible in our simulated \Chandra\ observations until destroyed at \taccrete.  The material forming this structure is not stripped from the central core but from the region just outside the initial cooling radius.  Hence, it is of intermediate entropy ($300-500$\keVcmsq) and temperature ($\sim 6.5$\keV).  No strong pressure gradients arise across it, as shown in the SZ maps in Fig. \ref{fig-1to1_Tsl}.

The bridges formed in our off-axis 3:1 mergers remain detectable for only $1.0$\Gyr\ following \tclosest, or roughly until \tapo.  As in the 1:1 cases, the entropy of the bridge material is of intermediate entropy ($200-400$\keVcmsq) and temperature ($\sim 4.5$\keV) and exhibits no strong pressure gradients.

In Figs. \ref{fig-1to1_Sigmagas} and \ref{fig-3to1_Sigmagas} we also see that the plumes generated from the disturbance of the primary and secondary cores are observable in the 1:1 cases but only marginally so in the 3:1 cases.  Figure \ref{fig-10to1_Sigmagas} shows that in the 10:1 cases, although the system appears disturbed after \tclosest, no extended features exist with sufficient surface brightness to be detected.

At \taccrete, very significant temperature fluctuations ($\Delta T/T\sim 20\%$) are present in all cases and persist until well after \trelax.  Because the accreting streams retain angular momentum from the merger and do not readily mix with the remnant, they produce spiral structures in the remnant's temperature and entropy distribution which are visible in our temperature and entropy maps at \trelax.

Because the SZ signature is proportional to gas density ($\rho_g$) rather than the $\rho_g^2$ dependence of the X-ray surface brightness, extended low density features tend to be washed-out in the SZ maps.  Despite this, we can see that our SZ maps indicate that the bridges formed in the 1:1 and 3:1 mergers remain above our nominal SZ threshold.  At \trelax~, when accreting streams leave the X-ray surface brightness elongated in these cases, the SZ maps look significantly more regular with only the 1:1 $v_t/V_c=0.15$ case looking comparably disturbed to its X-ray surface brightness. 

\subsection{Core oscillations}\label{analysis-oscillations}

In several of the cases we have studied, we find that the central ($r\lesssim0.5R_{500}$) ICM of our merger remnants oscillate with respect to their dark matter distributions. In the discussion of the evolution of disturbed morphology and relaxation which follows, there will be several instances when these oscillations will be relevant.  Such oscillations have been discussed by several authors in attempts to account for ``edges'' observed in the central surface brightness of some clusters (see Section \ref{analysis-edges} for more details and references).

In Fig. \ref{fig-rv_t} we present the displacement of the centre of mass of the densest central $150$\kpc\ of our remnant gaseous cores relative to the centre of mass of the densest central $150$\kpc\ of dark matter.  To differentiate oscillating modes from the influence of substructure passing through the core in this calculation, we also present the mass averaged velocity of the gas in the direction of this displacement's vector.  Displacement due to coherent oscillations will be accompanied by positive velocities until peak displacement when the velocity turns around and becomes negative, returning to zero when the displacement does.  The displacement of the core due to the arrival of substructure is revealed by increases in the displacement at times of small or negative radial velocity.  There are several instances of significant core oscillations following \taccrete\ stimulated by the second pericentric passage of the secondary core.  They generally last until \trelax.

We have performed this calculation for material within $50$\kpc\ as well.  On these scales we observe no oscillations larger than the softening length ($10$\kpc) at times when oscillations are present on $150$\kpc\ scales.  Hence, gas in the central regions of our remnant cores (where cooling is most rapidly taking place) remains tightly coupled to the dark matter potential with the surrounding material constituting the oscillation seen on $150$\kpc\ scales.  It has been hypothesised that energy from bulk oscillations in the potential of the core due to displacements of the gas from the dark matter could act as a source of heating for compact cool cores \citep{Markevitchetal01}.  Our analysis suggests that although oscillations may occur on $150$\kpc\ scales, the actively cooling gas on $50$\kpc\ scales can remain strongly coupled to the dark matter.  In such cases, oscillations in the potential may not be affecting cooling.  Further study is required to determine if this is the case for oscillations stimulated through close encounters by secondary systems with mass ratios of only a few percent, such as those studied by \citet{TittleyandHenriksen04}.

All of these calculations determine the motion of the gaseous core relative to the dark matter core.  We have checked to see how the dark matter core is itself moving within its halo.  For our 1:1 cases, we find that on both $150$ and $50$\kpc\ scales the dark matter cores remain stationary relative to the centre of mass of the system following \taccrete.  However, the dark matter cores of our 3:1 and 10:1 mergers experience significant motions about the system's centre of mass until well after \trelax.  The oscillations we observe between the gas and dark matter on $150$\kpc\ scales are thus a result of the dark matter core orbiting the centre of mass of the system.  As the gravitationally dominant dark matter core moves, the gaseous component is dragged behind it but with a lag due to the extra pressure it experiences.  The result is an oscillation between the two components as the dark matter core moves in its orbit.  On $50$\kpc\ scales, baryonic material dominates the potential of most of the gas and the dark matter can follow it since it is not subject to pressure.  As a result, oscillations between the gas and dark matter do not occur on these scales.

Following \taccrete, the gaseous and dark matter cores of our 1:1 merger remnants rapidly become concentric by \trelax.  A few displacements of the core position with respect to the dark matter are apparent but these are generally accompanied by negative radial velocities indicating that they are a product of the accretion of substructure onto the core.

The cores of our 3:1 merger remnants all experience bulk oscillations between their gaseous and dark matter components following \taccrete.  In the head-on and $v_t/V_c=0.15$ cases, the amplitude of these motions are $100$\kpc\ and $150$\kpc\ respectively with velocities reaching $400$\kms.  These motions last for $\sim 1$\Gyr\ and end immediately before \trelax.  In the $v_t/V_c=0.4$ case, a series of 4 core oscillations occur beginning at \taccrete\ with amplitudes as large as $50$\kpc\ and velocities of $175-250$\kms.  They are of much shorter duration ($\sim 0.5$\Gyr) and are finished by \trelax.

Some part of the dark matter cores of the secondary systems in our 10:1 mergers remain intact in near circular orbits within $50$\kpc\ of the remnant's centre until the end of our simulations.  This is not surprising since the sinking timescale due to dynamical friction is the longest.
As a result, we measure small amplitude ($<20$\kpc) oscillations in our 10:1 merger remnants.  These motions are most coherent in the head-on case and involve bulk velocities of $<100$\kms.

\section{Evolution of disturbed morphology}\label{analysis-disturbed}

\begin{figure*}
\begin{minipage}{175mm}
\begin{center}
\leavevmode \epsfysize=10cm \epsfbox{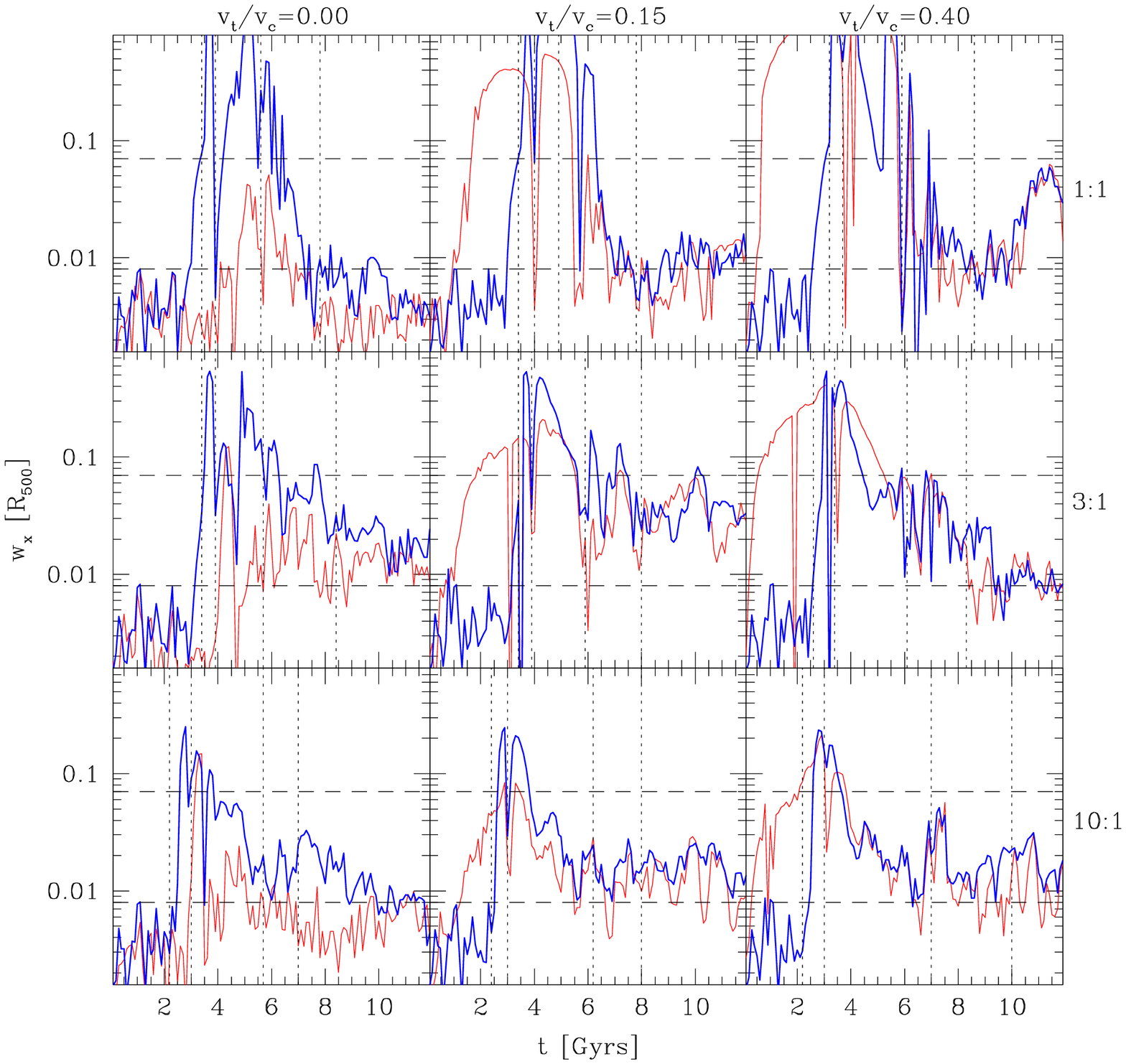}
\caption{Evolution of the centroid variance  computed from the X-ray surface brightness maps in the $x$ (line of sight) and $z$ (orthogonal) projections (thin red lines and thick blue lines respectively).  Horizontal dashed lines indicate the range of centroid shifts observed by \citet{OHaraetal05} while vertical dotted lines indicate (from left to right) $t_o$, \tclosest~, \taccrete\ and \trelax.  Centroids are computed within $1.2$\Mpc\ (roughly the initial $R_{500}$ of the primary).  The centroid statistic traces the qualitative evolution of the merging system very well.  The statistic takes on its maximum values when the system is in its most disturbed state and typically, it falls off between 
$t_{accrete}$ and $t_{relax}$ as the system recovers.   Over the course of the merger, the statistic spans the entire range of values found by \citet{OHaraetal05} in an analysis of 45 clusters$^\dagger$.}
\label{fig-wx_t}
\end{center}
\end{minipage}
\end{figure*}

With the general evolution of merging systems accounted for, we now discuss the timescales by which evidence of the merger event disappears.  Determination of the degree to which a cluster is disturbed has generally taken three approaches in observational studies: visual inspection, measurement of displacements of the X-ray peak from the X-ray or mass centroid of the system, and power-ratio analyses of the system's X-ray surface brightness.  In this section we shall apply each of these approaches in turn to establish when our simulated merger remnants would return to apparently undisturbed states.  In the section which follows, we shall compare these timescales to more formal and physical measures of the system's dynamical state.

\subsection{Simulated observations}\label{analysis-observations}

In evaluating the effectiveness of quantitative measures of our remnants' dynamical state, we shall use as a fiducial the time when our systems appear relaxed under reasonable observational circumstances.  We determine this time by visual judgement of our simulated \Chandra\ observations and of the uniformity of their isophotes; a process we facilitate by generating unsharp masked images with smoothing kernel sizes of $1.5$'' and $15$'' \citep{Chatzikosetal06}.  This is all done in projections normal to the plane of the orbit.  Substructure is likely most identifiable in this projection and thus, this measure represents a conservative upper limit on the time when the system would look disturbed.  We have studied how our results change if evaluation is performed in $x$-projections and find that the systems tend to look undisturbed as much as $2$\Gyrs\ earlier.

Generally speaking, evidence of substructure disappears first from the simulated images, then from the unsharp masked images and finally in the structure of the X-ray isophotes.  The delay between the dissolving of substructure in the images and the regularisation of their isophotes can take several \Gyrs\ and varies significantly from case to case.  Since the isophotes are the most sensitive measure, the system will be said to appear relaxed at \trelax\ when all isophotes brighter than $6\times$ the background level (\ie all but the faintest in Fig. \ref{fig-1to1_Sigmagas}, \ref{fig-3to1_Sigmagas} and \ref{fig-10to1_Sigmagas}) appear regular by visual inspection.  These times are tabulated in Table \ref{table-times}.  

In one case (1:1 head-on), minor complications arose during this procedure.  Although the simulated observations appear regular and relaxed by $t\sim 3.8$\Gyrs, their isophotes remain significantly boxy until $t\sim 4.4$\Gyrs.  Some evidence of this boxy morphology lingers until the end of the simulation making a clear determination of \trelax\ difficult.

\subsection{Centroid shifts}\label{analysis-centroids}

The apparent displacement of a system's core from the centre of the system has been used as a measure of dynamical disequilibrium by several authors and has been implemented in several different ways.  For instance, \citet{Mohretal93} quantify X-ray centroid shifts ($w_x$) using the offset of a system's X-ray surface brightness peak from its surface brightness centroid (integrated within a projected radius $r_p$).  They argue that systems for which $w_x$ is dependent on $r_p$ are dynamically young.  For a sample of 46 \textit{Einstein} observed clusters, they obtain an average value $\langle w_x \rangle=62$\kpc\ with $71$\% of systems having $w_x>0$ at a $3\sigma$ or greater significance.  These results generally conform with more recent results of \citet{Kolokotronisetal01} who find $\langle w_x \rangle=83$\kpc\ for a sample of 22 ROSAT observed systems.

We have chosen to quantify the centroid variance of our systems using the method employed by \citet{OHaraetal05}.  This approach determines the offset between the X-ray peak and centroid as a function of $r_p\lesssim R_{500}$ and assigns to $w_x$ the RMS of the resulting profile.  We determine this profile by computing the centroid within a circular aperture of size $r_p=R_{500}$ and then reducing its size by $5\%$ until it reaches $r_p\leq0.05R_{500}$.  We have experimented with various details of this procedure and find that it is essential to exclude the bright central core if the statistic is to detect subtle distortions.
Hence, we excise the central $r_p=30$\kpc\ when computing centroids (but retain it when we compute the position of the peak).

In Fig. \ref{fig-wx_t} we plot the time dependence of $w_x$ measured in this way for each of our simulations in $x$ and $z$-projections.  The evolution of this quantity in $y$-projections is very similar to its evolution in the $z$-projection and is omitted to clarify the plot.  We also indicate the range of typical values observed by \citet{OHaraetal05} in their study of 45 ROSAT observed clusters ($w_x=0.008R_{500}$ to $0.07R_{500}$, with 5 extreme points excluded) as well as the times of several important stages in each system's evolution.  

As a product of our idealised initial conditions, all of our systems start with values of $w_x$ less than the lower bound of the observations.  In off-axis cases, it immediately begins to increase in $x$-projections as the cores slowly move apart in projection.  This effect grows more significant with increasing $v_t/V_c$ as we would expect.  In $z$-projections, no increase occurs until $\sim t_o$ when the body of the  secondary enters the $R_{500}\sim1.2$\Mpc\ aperture used for our calculations.  
In the off-axis cases, the statistic for the $x$ and $z$-projections are virtually indistinguishable following \tclosest.  In the head-on cases, the signal is much lower when the merger proceeds along the line of sight.  Typically, the statistic drops sharply when the primary and the secondary are close together, as in the line-of-sight/head-on case and more generally at $t_{closest}$ and $t_{accrete}$.  This is due to the fact that the centroid shift in all the apertures is nearly the same.  In detecting disturbances, the centroid variance method depends on the displacement varying as the aperture is reduced and successively more distant surface brightness fluctuations are excluded from consideration.

\begin{figure*}
\begin{minipage}{175mm}
\begin{center}
\leavevmode \epsfysize=10cm \epsfbox{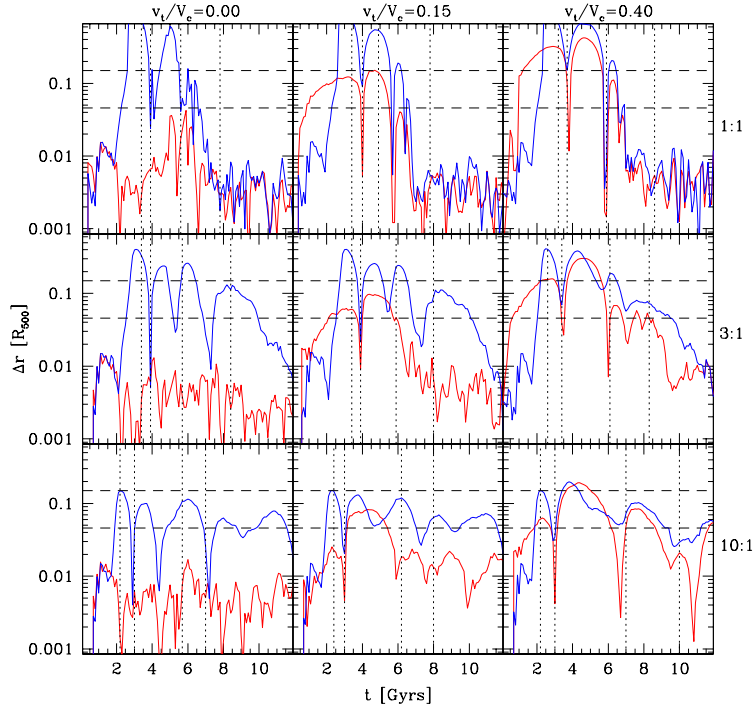}
\caption{Time evolution of the displacement between the X-ray peak and the centroid of the projected mass maps in the $x$ (line of sight) and $z$ (orthogonal) projections (thin red lines and thick blue lines respectively).  Horizontal dashed lines indicate the mean and maximum centroid shifts measured by \citet{Smithetal05} (normalised by $R_{500}$ of our system) while vertical dotted lines indicate (from left to right) $t_o$, \tclosest~, \taccrete\ and \trelax.  In the case of massive (1:1) mergers, the  statistic behaves like $w_x$ and tracks the morphological evolution of the system over the course of the merger.   For moderate and minor mergers, the statistic remains high well past $t_{relax}$ although the system is close to being virialised for $t > t_{relax}$.  This signal is entirely due to low-mass remnant dark matter substructure that keeps jiggling the mass centroid$^\dagger$.}
\label{fig-Dr_t}
\end{center}
\end{minipage}
\end{figure*}

Our 1:1 mergers evolve to values of $w_x$ much larger than those observed for a duration of $\sim 4$\Gyrs.  During this time, the system is a distinct binary and is not representative of any of the systems present in the \citet{OHaraetal05} catalogue.  Following \taccrete, $w_x$ declines rapidly and reaches a roughly constant level near the minimum observed value $\sim 1$\Gyr\ before \trelax.  In our $v_t/V_c=0.4$ case, there is a subsequent rise at late times due to a late accretion of material.  It is likely that the strength of this rise would not be as significant in a system hosting realistic substructure where the effect of late accreting material from multiple mergers would tend to cancel their individual effects.  For this reason, we will consider the system to have relaxed according to this measure $\sim 1$\Gyr\ prior to \trelax\ in this case but acknowledge that more study is required to understand clearly the role long lasting substructure has on $w_x$ in realistic environments.

In the 3:1 cases we see a similar rise after $t_o$ to values greater than the maximum observed but lasting for a much shorter duration ($\sim 2$\Gyrs).  Following \taccrete, $w_x$ oscillates erratically amongst the highest values observed by \citet{OHaraetal05}.  It slowly declines to the minimum observed value by the end of the simulations in the head-on and $v_t/V_c=0.4$ cases but remains high in the $v_t/V_c=0.15$ case.  The cause of this slowed or absent decline in $w_x$ following \taccrete\ is a sustained shift of the centroid in the central $0.5R_{500}$.  This is due to oscillations in the remnant's central dark matter distribution.

Lastly, our 10:1 cases are disturbed to high levels of $w_x$ for $\sim4$\Gyrs\ following $t_o$ and for $\sim1$\Gyr\ following \taccrete\ in the $v_t/V_c=0.4$ case (due to the reduced disruption of the core).  They oscillate just above the minimum observed value otherwise.

Thus, our systems appear exceptionally disturbed to our centroid variance measure for $\sim4$\Gyrs\ between \tclosest\ and \taccrete\ when they generally appear as two distinct clusters.  Immediately following \taccrete\ for our major/moderate (1:1 and 3:1) mergers and $\sim2$\Gyrs\ after \tclosest\ for our 10:1 mergers (when the systems appear as single remnants), our simulations all exhibit values of $w_x$ similar to the maximum observed by \citet{OHaraetal05} (with the natural exception of the head-on cases in $x$-projections).  In 3:1 cases, $w_x$ remains high until well past \trelax.  Our equal mass and 10:1 mergers quickly relax to levels comparable to the minimum values observed by \citet{OHaraetal05}.  Referring to Table \ref{table-times} we can see that for a cluster observed at $z=0.1$, most 10:1 mergers occurring before $z\sim0.5$ will have evolved past \taccrete\ and will thus yield $w_x\sim0.01R_{500}$.  This fact combined with the expectation that most clusters should have experienced such an event naturally explains the observed minimum of this statistic.

We have also computed this statistic for SZ maps ($w_{SZ}$) of our systems.  We find that the evolution of $w_{SZ}$ is nearly identical to that of $w_x$, with both producing very similar maximum shifts and periods during which $w_x$ is significantly higher than $0.1R_{500}$.

Another approach to measuring centroid shifts has been implemented by \citet{Smithetal05} who measure the shift between the peak in the X-ray surface brightness and the centre of projected mass of the system ($\Delta r$), as determined from weak lensing maps.  For a sample of 10 clusters, they find $\langle \Delta r \rangle=55$\kpc\ ($\sim0.045R_{500}$ for our systems ) with $70$\% of their systems having $\Delta r>0$ at a $3\sigma$ or greater significance.  In Fig. \ref{fig-Dr_t} we present the evolution of this statistic for our simulations.  Again, we plot only $x$ and $z$-projections due to the similarity of results in $y$ and $z$-projections.

Examining Fig. \ref{fig-Dr_t} we see that for the 1:1 mergers, this statistic drops precipitously after \taccrete\ (similarly to $w_x$) and sustains similar values after \trelax.  For the 3:1 mergers however, there is a long lived late time shift in $z$-projections which does not reach maximum amplitude until \trelax\ and does not reduce to $\Delta r\sim0.01R_{500}$ until shortly before the end of our simulations.  A similar but even longer lived shift persists in our 10:1 mergers which fail to sustain $\Delta r\lesssim 0.01R_{500}$ for more than $0.5$\Gyrs\ in any projection for any case.  These sustained shifts are due to central dark matter oscillations (see Section \ref{analysis-oscillations}).

Given the relatively high frequency of 10:1 mergers and the long duration of their impact on $\Delta r$, we expect that few systems should exhibit $\Delta r<0.045R_{500}$ (the mean value for our 10:1 $v_t/V_c=0.15$ merger after \trelax) and therefore, it is surprising that \citet{Smithetal05} find any systems consistent with $\Delta r=0$.  However, if projection effects are considered, it may be that some or all of the $3$ systems to which they assign relaxed morphologies using this statistic are being seen with their X-ray peak merely in projection against their centre of mass.  Alternatively, long lived substructure from multiple past minor mergers may compete against each other in their influence on the centre of mass, reducing its overall offset from the X-ray peak.  Further study of this statistic and its efficacy as an indicator of disturbed systems in a full cosmological context is clearly necessary.

\subsection{Power ratio morphology}

The use of surface brightness moment decomposition for quantifying the degree to which systems appear relaxed was first introduced by \citet{BuoteTsaiI} and has since been utilised in several observational \citep{BuoteTsaiII,Jeltemaetal05} and theoretical studies \citep{Thomasetal98,OHaraetal05}.  This method uses power ratios ($P_m/P_0$) computed from moments of the system's surface brightness distribution; $\Sigma(x)$, where $x'=(R,\phi)$ is in polar coordinates.  Following the procedure of \citet[][J05 hereafter]{Jeltemaetal05}, we have applied this procedure to our simulations computing power ratios from

\begin{center}
\begin{align}\label{eqn-power_ratios}
P_0(R)&=\left[a_o\ln\left(R\right)\right]^2\\
P_m(R)&=\frac{1}{2m^2R^{2m}}\left(a_m^2+b_m^2\right)
\end{align}
\end{center}

\noindent where $a_m$ and $b_m$ are moments of the distribution given by

\begin{center}
\begin{align}\label{eqn-moments}
a_m(R)&=\int_{R'\le R} \Sigma(x')(R')^m\cos m\phi'\,d^2x'\\
b_m(R)&=\int_{R'\le R} \Sigma(x')(R')^m\sin m\phi'\,d^2x'
\end{align}
\end{center}

\noindent These statistics are usually normalised by $P_0$ (\ie expressed as $P_m/P_0$), a measure of the total luminosity of the system.

We have studied the effects of removing the central cores of our simulated clusters when computing these statistics.  We find this to be particularly necessary when the system is only subtly disturbed.  At such times, the contribution of any persisting structure in $\Sigma(x)$ to Eqns. 4 and 5 becomes overwhelmed by the bright and centrally concentrated cores of the merging systems.  When we exclude the central cores from our calculations of $P_m$ we find a significant increase in amplitude and scatter.  For the discussion which follows, we have computed power ratios with the central $30\kpc\sim0.025R_{500}$ excised.  We have confirmed that there is little effect on our conclusions if this radius is reduced to $15$\kpc\ or increased to $50$\kpc.

\begin{figure*}
\begin{minipage}{175mm}
\begin{center}
\leavevmode \epsfysize=10cm \epsfbox{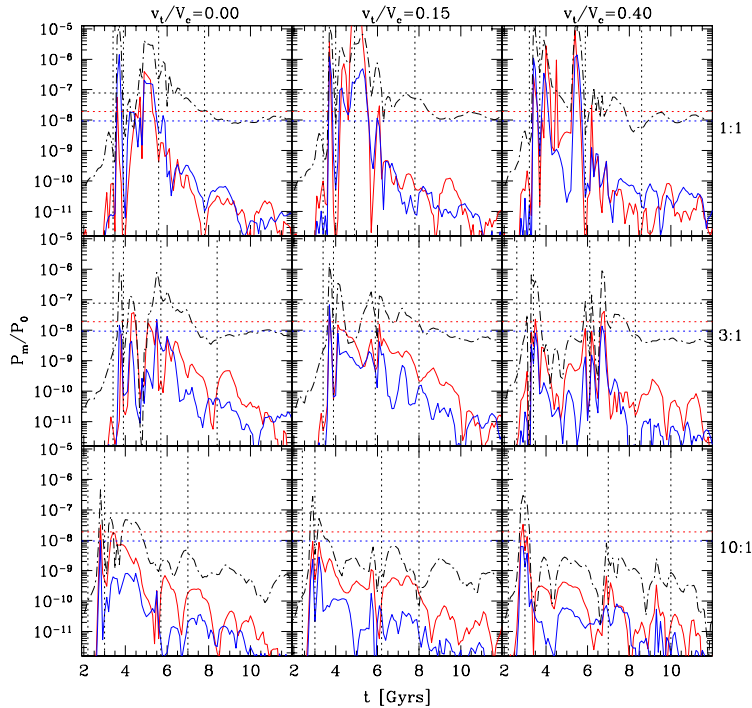}
\caption{The time evolution of the X-ray surface brightness power ratios computed within a $0.5$\Mpc\ aperture in the $z$-projection, the
projection in which the merger-induced disturbances are the most apparent.  Dotted horizontal lines depict the median noise levels for the 
observational sample analysed by \citet{Jeltemaetal05} for (in descending order) $P_2/P_0$, $P_3/P_0$ and $P_4/P_0$.  Dash-dotted black, thick
red and blue lines correspond to $P_2/P_0$, $P_3/P_0$ and $P_4/P_0$ respectively.  Vertical dotted lines represent (from left to right) $t_o$,
\tclosest, \taccrete\ and \trelax.  The power ratios can easily pick out the very short-duration highly disturbed states but for the most part
the amplitudes tend to be comparable to or less than the median noise levels for the J05 sample of clusters$^\dagger$.  See also Fig. \ref{fig-P4_P3}.}
\label{fig-PN_t}
\end{center}
\end{minipage}
\end{figure*}

In Fig. \ref{fig-PN_t} we plot the time evolution of $P_2/P_0$, $P_3/P_0$ and $P_4/P_0$ in a $0.5$\Mpc\ $z$-projected aperture for each of our
mergers.  \citet{BuoteTsaiI} find that model systems with power ratios identically equal to zero can be assigned significant values (a few 
times $10^{-7}$ for $P_3/P_0$ for instance) when realistic noise is added to simulated X-ray images.  Based on their analysis of an observed 
sample of clusters, J05 found that median noise contributions resulted in values for $P_2/P_0$, $P_3/P_0$ and $P_4/P_0$ of $7.5\times10^{-8}$, 
$1.9\times10^{-8}$, and $9.4\times10^{-9}$ respectively.   We indicate these levels in the plot.  The plot illustrates that in the absence of 
noise, the values of the power ratios for the mergers are typically well below the median noise levels, except when the systems are in their 
most disturbed state.  In practise, the values assigned to a relatively relaxed-looking cluster will be highly dependent on the quality of the 
individual observation.  In fact, the lowest values of $P_N/P_0$ quoted by J05 are a factor of $\sim 10$ below the median noise levels.  
In what follows we shall loosely refer to the median noise levels reported by J05 as the ``observable levels'' of each 
statistic.  We will treat these levels as a reasonable fiducial when determining the point at which our systems would be judged as undisturbed 
by this approach.  


The evolution of $P_3/P_0$ and $P_4/P_0$ is very similar; the latter behaves more or less like a scaled version of the former.  In the approach
to the first pericentric passage at \tclosest, both power ratios experience a brief sharp spike which exceeds observable levels in all cases, though just barely in the 3:1 and 10:1 mergers.
This is followed by a climb to a plateau or a broad peak, during which time the signal remains at or below the observable levels except in the 1:1 case, and a fall-off at $\sim t_{accrete}\pm 1$\Gyr.  The amplitudes of the initial spike and of the plateau/broad peak are correlated with the secondary's mass, with the most massive merger generating the greatest signal.   There does not appear to be any clearly discernible correlations with impact parameter.  Once the signal falls below observable levels after the second drop, it remains so for the duration of the simulation under all conditions.

The evolution of $P_2/P_0$ is  different from that of $P_3/P_0$ and $P_4/P_0$.  Like the latter two, the signal briefly spikes above the noise level at \tclosest~but rapidly drops below. It exceeds the noise level again at \taccrete~for $\lesssim 2$\Gyr~and then generally only for the mergers with mass ratios greater than 3:1.  (At 3:1, the second spike just exceeds the observable level).
Afterwards, it stabilises at values far below observable levels, changing very little after \trelax.  For the major/moderate (1:1 and 3:1) mergers,  $P_2/P_0$ settles at $\sim10^{-8}$ while our 10:1 mergers, the power ratio hovers at values of about $\sim10^{-9}$.

In their study of the redshift evolution of power-ratios, J05 find that between $z=0.5-0.9$ (their high-$z$ sample) and $z<0.5$ (their low-$z$ sample), the $P_3/P_0$ and $P_4/P_0$ ratios show statistically significant declines while the $P_2/P_0$ ratio shows no significant change.  Furthermore, observed values of $P_2/P_0$ are confined to a range more narrow than $P_3/P_0$ or $P_4/P_0$.  Our simulations reproduce these trends but with a much lower normalisation.  Further study is required to understand if this discrepancy is a product of noise effects or our idealised initial conditions and lack of cosmological substructure.  Since $P_2/P_0$ is the power ratio most sensitive to the large scale ellipticity of the system, it would be interesting to determine if more realistic triaxial initial conditions would produce a similar narrow range of near constant final $P_2/P_0$ but with a higher normalisation comparable to that observed.  As it stands, our results suggest that the power ratios are most sensitive to gross disturbances such as those that arise early in the merging process.

\begin{figure*}
\begin{minipage}{175mm}
\begin{center}
\leavevmode \epsfysize=10cm \epsfbox{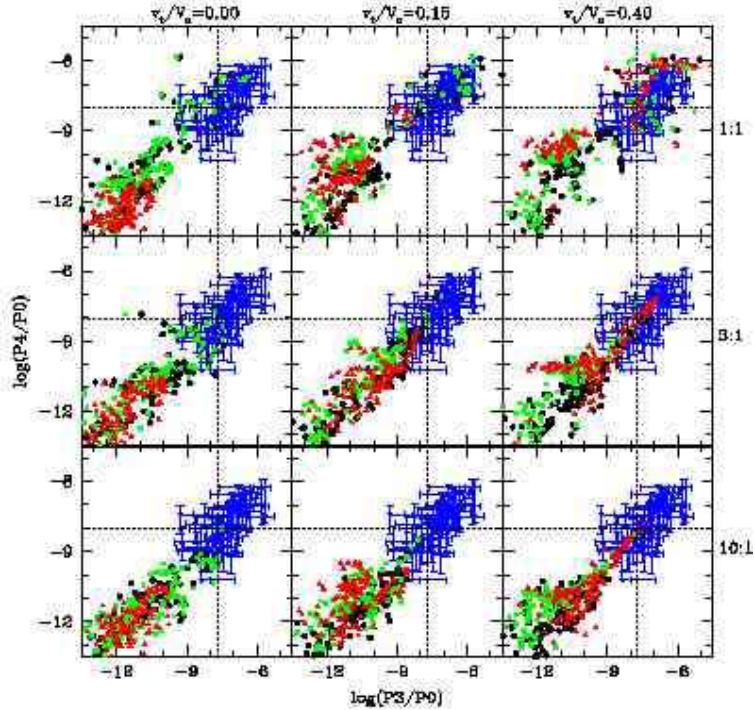}
\caption{Plot of $P_4/P_0$ vs. $P_3/P_0$ for our simulations measured at different time snapshots over the course of the merger, within an aperture of $0.5$\Mpc.  
The data of \citet{Jeltemaetal05} is plotted in blue with our simulations plotted as red triangles, green squares and black circles for $x$, $y$
and $z$ projections respectively.  This combination of power ratios yields a signal above the median noise level of J05 sample (shown as
horizontal and vertical lines) only when system is caught in the short-lived highly disturbed state.   It is most sensitive to disturbances induced by highly off-axis mergers and least sensitive to line-of-sight mergers$^\dagger$.}
\label{fig-P4_P3}
\end{center}
\end{minipage}
\end{figure*}


Correlations between different power ratios have generally been used to identify disturbed systems and to study the evolution of cluster substructure.  Here, we focus our attention on the $P_4/P_0$-$P_3/P_0$ plane.  In Fig. \ref{fig-P4_P3}, 
we present a comparison in this plane between the observations of J05 and our simulations computed within apertures of $0.5$\Mpc. 
Results computed in $x$, $y$ and $z$-projections are presented with red triangles, green squares and black circles respectively.  In Fig. \ref{fig-P4_P3}, we can see that the observed trends in 
normalisation, maximum amplitude and scatter are well reproduced, though our measurements cluster at values of $P_3/P_0$ and $P_4/P_0$ well below the observations.  These two power ratios return values 
above the observable levels only during the highly disturbed phases of the merger, and  during the initial stages of the off-axis mergers.

We have also examined the distribution of points in the $P_4/P_0$-$P_3/P_0$ plane, where the power ratios are computed within a larger aperture of radius $R_{500}\sim 1.2$\Mpc.  With the exception of some reduction in the scatter, the results are very similar to those presented in Fig. \ref{fig-P4_P3}


Finally, we also applied the power ratio analysis to total projected mass and SZ maps to examine the efficacy of this approach to measuring substructure through weak lensing and microwave observations.  
The results were indistinguishable from the X-ray results, with amplitudes of the $P_3/P_0$ and $P_4/P_0$ for the projected mass maps being lower.

\subsection{Comparison of disturbed morphology measures}

We have examined several methods of determining the dynamical state of galaxy clusters.  In Figs. \ref{fig-wx_t}, \ref{fig-Dr_t} and \ref{fig-PN_t} we present the time evolution of $w_x$, $\Delta r$ and $P_m/P_0$.  In each case we indicate the moment when the isophotes of our simulated $50$\ks\ $z=0.1$ \Chandra\ observations appear undisturbed as a fiducial reference for the sensitivity of each approach.

We find that while the centroid shift and power ratio approaches are both capable of identifying highly disturbed systems or systems with significant, well-defined substructures, the $w_x$ is more sensitive to subtler disturbances and its time evolution tracks the visual appearance as well as the dynamical state of the merging system quite well.  The amplitude of the power ratios is relatively low when the system is not highly perturbed (for example, the analysis barely registers 10:1 mergers) and appears to be  highly susceptible to noise in the X-ray images.  More study is required to understand in exactly what way realistic noise affects the sensitivity of the $P_m/P_0$ statistics.  

\begin{figure*}
\begin{minipage}{175mm}
\begin{center}
\leavevmode \epsfysize=10cm \epsfbox{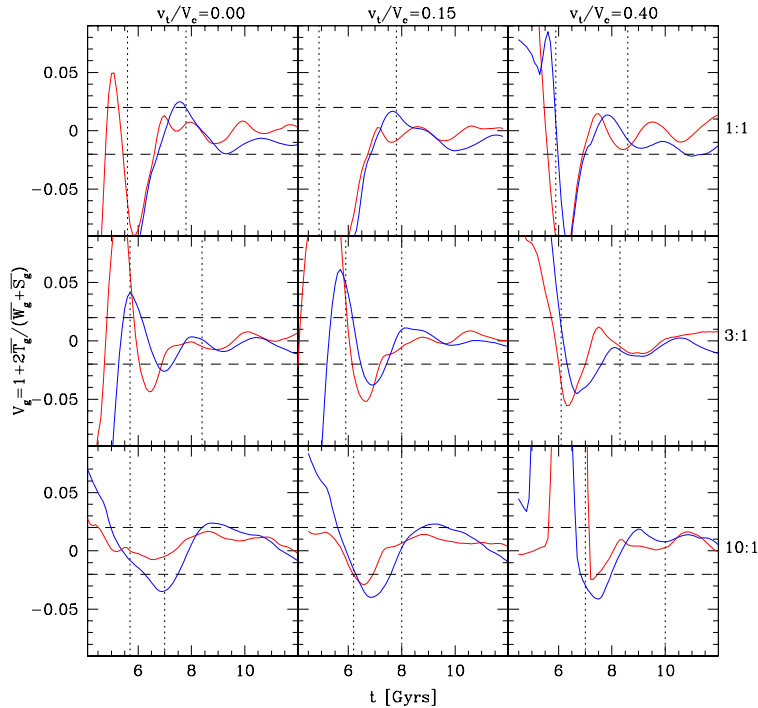}
\caption{Time evolution of the virial parameter computed over a volume of radius $R_{500}$ (thick red) and $R_{200}$ (blue) for the gaseous component of our simulations.  The evolution of the {\it gas} virial parameters for these two volumes is very similar.  The horizontal dashed lines indicate our $2$\% virialisation criteria.  The vertical dotted lines indicate \taccrete\ and \trelax\ for each case. If we define the system to be relaxed when the virial parameter drops to and remains at $\left| V \right| < 0.02$, then the gas component typically relaxes halfway between \taccrete and \trelax$^\dagger$.}
\label{fig-Vg_t}
\end{center}
\end{minipage}
\end{figure*}

Our experiments with the $\Delta r$ statistic introduced by \citet{Smithetal05} suggest that it is very sensitive to long lived substructures and oscillating modes in the dark matter haloes of merger remnants.  The statistic yields a signal even when dynamically the system is within 2\% of being relaxed
as defined by our virial indicator (see Section \ref{analysis-relaxation} and Fig.~\ref{fig-Vg_t}).
However, it is likely that our idealised mergers are exaggerating this sensitivity.  More study with cosmological simulations which include a proper account of halo substructure should be conducted to determine exactly how effective $\Delta r$ is relative to centroid shift and power ratio measures.

In summary, for clusters at $z=0.1$, the eye is remarkably adept at identifying disturbed morphologies in deep \Chandra\ images but quantitatively, we find that the centroid variance, $w_x$, is the most effective measure of a system's dynamical state.  

\section{Relaxation}\label{analysis-relaxation}

Is a system which appears undisturbed actually in a relaxed state?  This is commonly assumed to be the case in studies which have selected compact cool core systems with regular and symmetric isophotes to represent equilibrium systems.  However, all clusters are a product of mergers and even though their dynamical timescales are short ($\sim1.5$\Gyr\ at $R_{200}$), the timescales by which bound substructures become disrupted and dissolve into the accreting halo can be much longer, as illustrated in the preceding sections.
As a result, a merger remnant can take significant lengths of time to reach a proper equilibrium state, perhaps even longer than the typical interval separating merger events.

In practical terms, the point at which a system appears undisturbed depends on the state of the system's baryonic component whose signatures of disturbance can be diluted by observational limitations.  Furthermore, the gaseous component is subject to disruptive gas dynamical forces which do not act on the dynamically dominant dark matter component, perhaps erasing evidence of substructure faster than it loses dynamical relevance.  Hence, it is natural to suspect a discrepancy between when a system appears relaxed and when it is formally so.

In this section we will quantify the relaxation of our merger remnants by studying their recovery towards a virialised and hydrostatic state of equilibrium.  We shall compare the timescales by which they do this to the point at which the system would look relaxed under typical observational circumstances.  We shall find that our systems typically achieve a virialised state approximately when they appear relaxed but continue to exhibit deviations from hydrostatic equilibrium at the level of $10-20$\% until the end of our simulations.

\subsection{Virialisation}

To study the virialisation of our systems we employ the scalar virial theorem \citep[see][for good reviews]{Collins78,Spitzer78,BandT} which states that
\begin{center}
\begin{equation}\label{eqn-scalar_virial}
\frac{1}{2}\frac{d^2I}{dt^2}=2T+W+S
\end{equation}
\end{center}

\noindent where $T=K+U$ is the sum of the system's kinetic ($K$) and thermal ($U$) energies,
$W=\sum_{i=1}^{N} \vec{F_{i}} \cdot \vec{r_{i}}$ is the `virial', and 
$I$ is the system's moment of inertia, all computed within a specified volume.  The term denoted by $S=S_g+S_d$ is a surface pressure term integrated over the bounding surface of the volume, with contributions $S_g$ and $S_d$ due to the gaseous and collisionless (dark matter and stellar) components given by
\begin{center}
\begin{align}\label{eqn-surface_gas}
S_g&=-\oint P r\,dA\\
S_d&=-\oint \rho_d \sigma_d^2 \,dA
\end{align}
\end{center}

\begin{figure*}
\begin{minipage}{175mm}
\begin{center}
\leavevmode \epsfysize=10cm \epsfbox{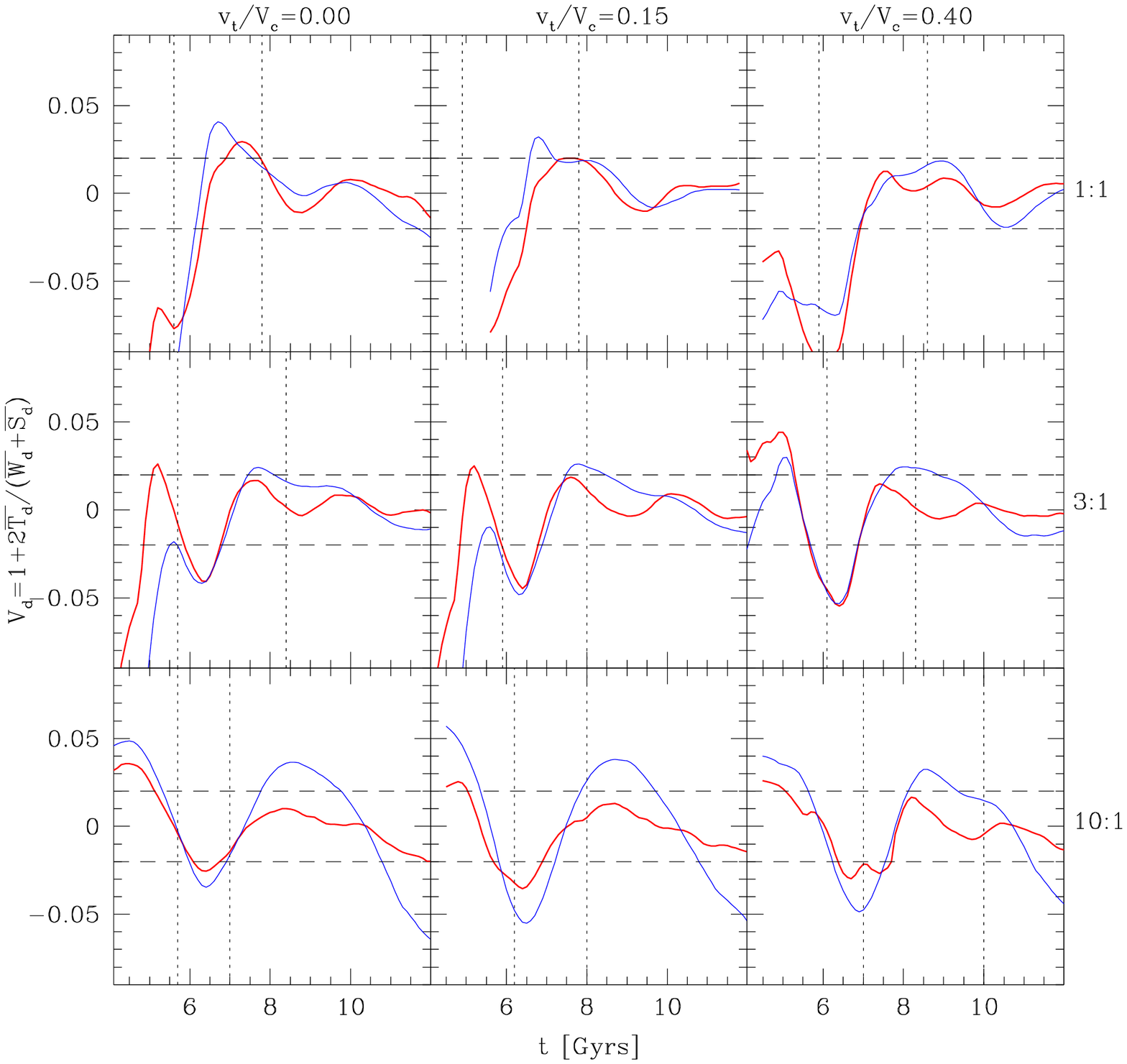}
\caption{Same as Fig. \ref{fig-Vg_t} but depicting the virialisation of the dark matter component of our simulations. The dark matter component typically relaxes slightly later than the gas component and the oscillations in $V$ damp out more slowly.  Also, the central regions of the systems relax faster than the full cluster$^\dagger$.}
\label{fig-Vd_t}
\end{center}
\end{minipage}
\end{figure*}

\noindent where $P=nkT$ is the pressure of the gas with a number density $n$ and temperature $T$ and $\sigma_d$ and $\rho_d$ denote the velocity dispersion normal to the surface and mass density of the collisionless component.



Averaged over an interval in which the system is periodic (or instantaneously for a steady state system), the left side of Eqn. \ref{eqn-scalar_virial} vanishes.  Furthermore, when integrated over the entirety of an isolated system, $S$ becomes zero and Eqn. \ref{eqn-scalar_virial} takes its popular form 
\begin{center}
\begin{equation}\label{eqn-virial_common}
2T+W=0.
\end{equation}
\end{center}


It is commonly taken for granted that $W$ represents the gravitational binding energy of the system but it must be emphasised that this is strictly true only when computed over the totality of an isolated system.  In general, $W$ is not equivalent to the gravitational binding energy within $r$.  

In Figs. \ref{fig-Vg_t} and \ref{fig-Vd_t} we present the temporal dependence of the virial disequilibrium parameter \citep{Ricker98} for the gaseous and dark matter components of our evolving merger remnants following \taccrete\ (when the system can be considered to be a single remnant).  
(Although the baryonic component's contribution to the virial parameter of the total system is negligible, it is interesting to see if the observable gaseous component virialises on the same timescales as the dark matter component.)  This virial parameter is defined as follows:
\begin{center}
\begin{equation}\label{eqn-virial_parameter}
V= 1+\frac{2\bar{T}}{\bar{W}+\bar{S}}
\end{equation}
\end{center}

\noindent where $\bar{T}$, $\bar{W}$ and $\bar{S}$ are time averages computed to account for the fact that the canonical form of the virial theorem presented in Eqn. \ref{eqn-virial_common} holds only when the quantities in Eqn. \ref{eqn-scalar_virial} are averaged over an interval in which the system is periodic.  This is done using a dynamical time $t_{dyn}=\sqrt{3 \pi / 16 G \Delta \rho_c}$ (where $\Delta$ is the over density being considered) for quantities involving the collisionless component and the local isothermal sound crossing time $t_{sc}=r\sqrt{\mu m_p/kT}$ for quantities involving the gaseous component.  Although the system may not be strictly periodic over these timescales, we have confirmed that the term involving $I$ in Eqn. \ref{eqn-scalar_virial} contributes negligibly during the period we present in figures \ref{fig-Vg_t} and \ref{fig-Vd_t}.  Following RS01, we select the moment of virialisation to be when the system sustains a value of $\left| V \right| < 0.02$.

Overall, we find that the virial parameters for the gas and the dark matter track each other's general trends although the oscillations in the gas virial parameter tend to be of slightly lower amplitude and damp more quickly.  As a result, the gas component typically relaxes (\ie the gas virial parameter drops to and remains at $\left| V \right| < 0.02$) earlier than the dark matter component. There is not much difference in the relaxation time of gas within $R_{500}$ and $R_{200}$.  On the other hand, the dark matter distribution within $R_{500}$ relaxes earlier than the full cluster.

In more detail, our 1:1 mergers all virialise by \trelax.  The gaseous component generally virialises $\sim 0.5$ \Gyrs\ before the dark matter component which virialises almost exactly at \trelax\ in all cases.  Thus, the system is virialised approximately $4.5$-$5.5$\Gyr\ (or $3$-$4$ dynamical times) after the secondary initially crosses $R_{200}$ of the primary \citep[in good agreement the findings of previous studies, \eg~][]{Roettigeretal97}.  The virialisation of our 3:1 mergers proceeds similarly to our 1:1 mergers for the dark matter. 
In the 10:1 case, the gaseous component within $R_{500}$ achieve virialisation at or well before our 1:1 and 3:1 mergers: immediately after \taccrete\ in the off-axis cases and well before in the head-on case.  Within $R_{200}$, however, our 10:1 gaseous components reach virialisation much later ($\sim 2$\Gyrs) than our 1:1 and 3:1 cases.  The dark matter behaves similarly.  Within $R_{500}$ the system is virialised at or before \trelax~while within $R_{200}$, the system never reaches our virialisation criteria.  In these cases, the system sustains a state where $V_d$ oscillates with an amplitude of $\sim 5$\% until the end of our simulations.  The retarded onset or absence of virialisation within $R_{200}$ for our 10:1 mergers is a result of late infall of material dispersed beyond $R_{200}$ during the merger.

\begin{figure*}
\begin{minipage}{175mm}
\begin{center}
\leavevmode \epsfysize=10cm \epsfbox{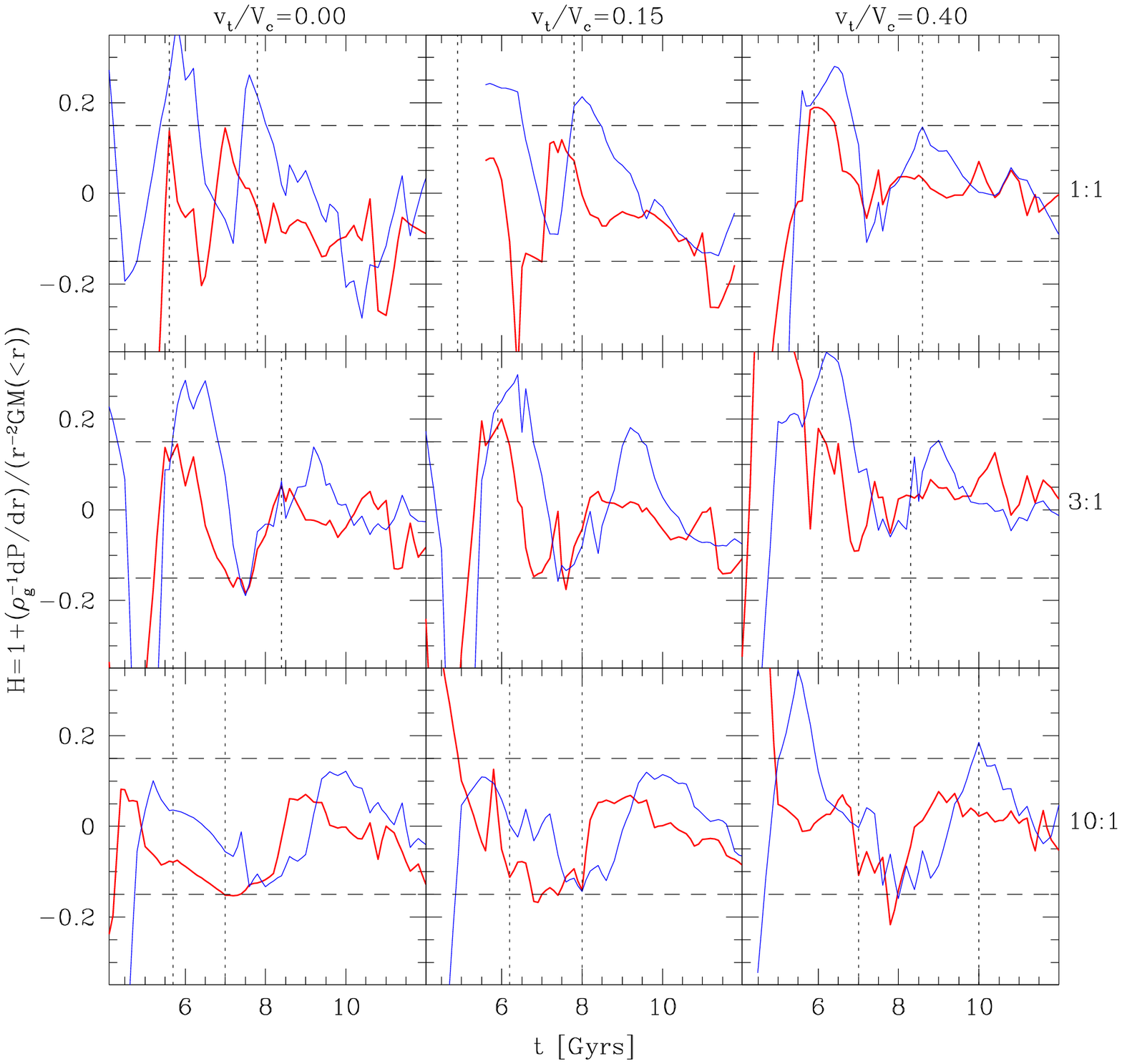}
\caption{Hydrostatic disequilibrium parameter measured at $R_{500}$ (thick red) and $R_{200}$ (blue).  The horizontal dashed lines indicate $15$\% equilibrium (the criteria used for $t_{hydro}$ in Table \ref{table-times}).  The vertical dotted lines indicate \taccrete\ and \trelax\ for each case$^\dagger$.}\label{fig-HE_t}
\end{center}
\end{minipage}
\end{figure*}

\subsection{Hydrostatic equilibrium}

Another independent measure of the equilibrium status of our merger remnants is the dynamical criterion of hydrostatic equilibrium.  For a pressure supported medium in a gravitational potential, hydrostatic equilibrium implies
\begin{center}
\begin{equation}\label{eqn-hydrostatic_equilibrium}
\rho^{-1}\frac{dP}{dr}=\frac{G M(<r)}{r^2}
\end{equation}
\end{center}

Given that X-ray derived mass determinations of clusters assume that this condition holds, its validity following mergers is of significant importance.  Recent studies have found evidence suggesting that clusters for which hydrostatic equilibrium can be accurately assumed are rare but no consideration has yet been given to whether systems can fail to be in hydrostatic equilibrium while appearing undisturbed.  In Fig. \ref{fig-HE_t} we plot the evolving hydrostatic disequilibrium parameter of our merger remnants following \taccrete\ given by
\begin{center}
\begin{equation}\label{eqn-H_parameter}
H=1+\frac{\rho^{-1}\frac{dP}{dr}}{r^{-2} G M(<r)}
\end{equation}
\end{center}

The evolution of hydrostatic disequilibrium for our systems all follow the same qualitative pattern.  Shortly before \taccrete, at both $R_{500}$ and $R_{200}$, they experience significant positive excursions in $H$ as the secondary core accretes to the remnant core.  Following this, $H$ is seen to oscillate with damped amplitude in each case with the oscillations at $R_{200}$ lagging in phase behind those at $R_{500}$ by $\sim 0.5$\Gyrs.  The phase of these oscillations is generally such that at $R_{200}$, a period of positive $H$ lasting $\sim 2$\Gyrs with an amplitude of $10-20$\%, occurs at or shortly after \taccrete.  At $R_{500}$, slightly lower amplitude (typically $\sim 10-20$\%) and less coherent oscillations persist until the end of our simulations.

Regular disruptions of hydrostatic equilibrium such as those depicted in Fig. \ref{fig-HE_t} could translate into discrepancies in X-ray mass measurements obtained from methods which assume the validity of Eqn. \ref{eqn-hydrostatic_equilibrium}.  We shall examine this issue in more detail in \citet{P06b} where we will discuss the effects of mergers on global cluster observables and scaling relations.

\subsection{Comparison of apparent and formal states of relaxation}

In Section \ref{analysis-disturbed} we found that our merger remnants generally appear relaxed through centroid (our preferred method) and power ratio measures once the isophotes of our simulated $z=0.1$ \Chandra\ observations appear undisturbed at \trelax.

In Table \ref{table-times} we list the times at which our systems are formally virialised ($t_{virial}$; measured for the dark matter) or in hydrostatic equilibrium ($t_{hydro}$; taken to be when the system sustains $\left| H\right| <0.15$) at $R_{500}$.  Our systems are generally virialised (both the gas and dark matter) within $R_{500}$ at or near \trelax\ while late accretion of material renders the system somewhat less virialised ($5$\% versus our $2$\% virialisation criterion) at $R_{200}$ in our 10:1 mergers at \trelax.  We also list the redshifts at which the secondary would have to cross $R_{200}$ of the primary in order to be seen in each phase at redshifts of $z=0.1$ and $z=0.5$.  We can see from this how unlikely virialised or hydrostatic clusters should be at $z=0.5$.

We thus find that our merger remnants are generally virialised at or shortly ($\lesssim 2$\Gyrs) after appearing relaxed to visual inspection.  However, even at this time merger remnants are adiabatically oscillating, exchanging energy between thermal and kinetic components.  They are in hydrostatic equilibrium only at the $10$-$20$\% level at this time and sustain this level of hydrostatic disequilibrium with little change after appearing relaxed.

\section{Transient Structures}\label{analysis-transients}

Our simulations reveal that the phenomena generally referred to as ``cold fronts'' consist of several classes of transient structure formed by the disruption of compact cool cores in off-axis mergers.  Previous theoretical studies of cold-front production have focused primarily on the disruption of the secondary's core but our analysis reveals that both cores can be involved.  Several other transient phenomena such as core surface brightness discontinuities and oscillations are naturally formed in the process as well.  In this section we shall discuss in more detail the processes driving the formation of these structures, noting where possible the occurrence of each in presently published observations of cluster mergers.

\subsection{Bridges}\label{analysis-bridges}

At two points during a cluster merger, luminous bridges connecting the interacting systems can be formed.  In the first, the outer atmospheres of the merging clusters become compressed on the incident side of the collision during preinteraction.  The resulting increase in gas density and temperature between the systems leads to enhanced surface brightness between the two systems and the formation of a luminous bridge.  This effect is short-lived ($\sim 0.5$\Gyrs\ in the 1:1 cases, less in others) but has been observed in several systems including A399/401 \citep{Sakelliouetal04} and A1750 \citep{Belsoleetal05}.

\begin{figure*}
\begin{minipage}{175mm}
\begin{center}
\leavevmode \epsfysize=8cm \epsfbox{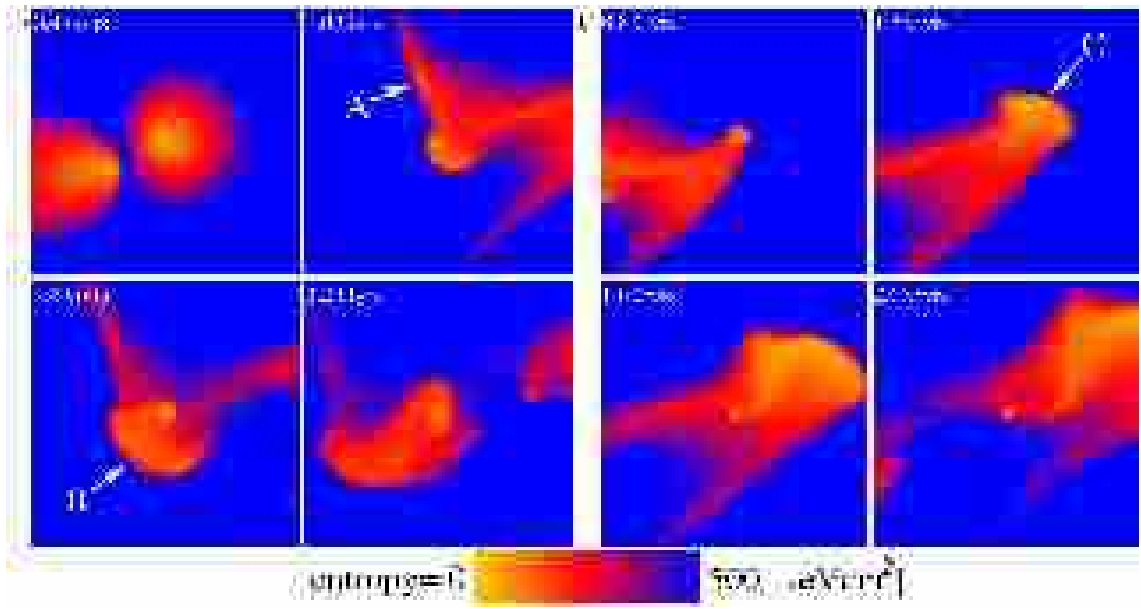}
\caption{The left panel of four plots shows entropy maps for our 3:1 $v_t/V_c=0.15$ simulation illustrating the ejection of two distinct
``plumes'' (labelled A and B) of low entropy material from the primary core in response to the pericentric passage of the secondary.   The 
first is triggered by the shocks that precede the secondary, and the second by negative pressure gradients in the wake behind the secondary.
The right panel of plots illustrates the ejection of a plume (labelled C) from the secondary at apocentre in the same simulation.  The plumes
eventually evolve into streams and most of the ejected low entropy material flows along the streams into the primary core.  
Each frame above  is $1.5$\Mpc\ on a side and is a $z$-projection of a $0.5$\Mpc\ thick slice$^\dagger$.}
\label{fig-3to1_plumes}
\end{center}
\end{minipage}
\end{figure*} 

As discussed in Section \ref{analysis-off-axis}, material stripped by ram pressure from the interacting clusters becomes dispersed, forming comet-like tails which trail their surviving cores following \tclosest.  Due to the density gradients of the systems they are passing through and the scaling of ram pressure with the density of the flowing material, this process is more efficient on the side of each core incident to the collision.  As a result, overlapping enhancements of stripped material form behind each core, producing a luminous bridge which joins the two systems.

Following \tclosest,  material from these regions moving in the plane of the orbit is driven together.  It becomes particularly concentrated in a curved planar feature oriented normal to the plane of the orbit.  As a result, the surface brightness of this bridge is most bright and concentrated in $z$-projections.  In $x$-projections it is not discernible and in $y$-projections it is significantly wider and more diffuse.  The cores of both clusters appear disturbed at these times, distinguishing this process from the one acting during preinteraction.

The dependence of the brightness and shape of the bridge could be used as a useful cue when interpreting the orientations of observed merging systems.  In A1758 \citep{DavidKempner04} for instance we see an excellent example of a system apparently exhibiting a post-interaction bridge.  The two merging systems in this case (A1758N and A1758S) both have disturbed cores and there is an obvious bridge connecting them.  The width of the bridge is more than $1.2$\Mpc\ suggesting that the system may be significantly inclined to the plane of the sky.  Other systems exhibiting similar morphologies include A1644 \citep{Reiprichetal04}, A754 \citep{Henryetal04} and A115 \citep{GutierrezKrawczynski05}.

\subsection{Plumes}\label{analysis-plumes}

In Section \ref{analysis-off-axis} we noted that in all of the off-axis cases we have studied, outwardly propagating collimated plumes of primary and secondary core material are produced between \tclosest\ and \tapo.  Several mechanisms are involved in the production of these structures, a few of which have been discussed previously in the literature.  \citet{Mathisetal05} provide a good summary of theoretical research involving the disruption of cold cores.  Their work supports the claims of \citet{Heinzetal03} who illustrate the important influence of ram pressure stripping during the disruption of merging cluster cores.  However, they emphasise that other processes are likely essential to the formation of cold fronts.  

Our study confirms and expands upon this claim.  Although ram pressure is certainly important for the disruption of both clusters, it is primarily only responsible for the stripping of material roughly beyond $R_{2500}\sim 500$\kpc\ (moderately high entropy material).  It does not lead to the prominent low entropy radial plumes produced in our simulations.  In our simulated \Chandra\ images, the period during which ram pressure stripped material is bright enough to produce an observable ``comet-like'' surface brightness feature is brief ($\sim 0.5$\Gyrs) and occurs only at or shortly after \tclosest.  A disturbed merging system spends most of its time at or near apocentric passage.  In our simulations, material stripped by ram pressure has dispersed and is no longer apparent in surface brightness by that point.  Bright features of displaced core material are produced primarily by other mechanisms during this longer lived phase.

In Fig. \ref{fig-3to1_plumes} we illustrate the production of three low entropy plumes generated from the cores of the primary (Fig. \ref{fig-3to1_plumes}a) and secondary (Fig. \ref{fig-3to1_plumes}b) with a series of entropy maps covering the evolution of our 3:1 $v_t/V_c=0.15$ simulation during first pericentric passage (Fig. \ref{fig-3to1_plumes}a) and apocentric passage (Fig. \ref{fig-3to1_plumes}b).  In these figures, the secondary enters from the bottom left and passes underneath and near to the core of the primary.

In Fig. \ref{fig-3to1_plumes}a we see that two plume-like structures (labelled ``A'' and ``B'') are produced from the primary core.  The one labelled ``A'' is produced on the side roughly opposite from the point of closest contact with the secondary core at \tclosest.  This feature is generated from material which has become confined in a narrow structure by the convergence of the shock as it passes the primary core.  This behaviour is produced by refraction of the shock due to the decline in the primary core's sound speed into the centre.

The other two plumes (``B'' and 'C' in Fig. \ref{fig-3to1_plumes}) are produced through distinctly different means.  In the off-axis simulations we have studied, short-lived but strong negative radial pressure gradients can be generated near the cores of the interacting clusters, leading to the ejection of low entropy material into the outer regions of the merger remnant.  Since these gradients are often very localised, the ejection tends to be highly directional leading to the formation of large outwardly flowing collimated plumes.  Since the process is adiabatic, the ejected material retains its low entropy and eventually reaccretes to the core of the remnant.  This occurs as low entropy streams of high velocity material over an extended period lasting well after \taccrete\ (see Section \ref{analysis-streams}).

Two mechanisms generally conspire to make this happen: the sudden release of compressive forces due to changes in the relative radial motion of the systems and the presence of low pressure regions in the direction toward which material is dispersed as a result.  After pericentric passage, the primary's core is rapidly compressed and the wake of the secondary's motion creates a large low pressure region on the incident side of the collision.  Low entropy material from the primary's core becomes ejected back along the secondary's path as a result, forming plume ``B''' which reaches a radius of $\sim 800$\kpc\ in this case.  

The secondary core is subject to these mechanisms at \tapo\ when it turns around and returns to the primary core.  The decline of ram pressure at that time results in the release of compressive forces and the strong negative pressure gradient of the primary at such radii provides a low pressure region for the secondary core's material to disperse towards.  These mechanisms are assisted by a gravitational ``sling-shot'' effect resulting from the lag of the secondary's gas core behind its dark matter halo, as noted by \citet{Bialeketal02}.  This lag is due to the additional pressure experienced by the gas but not the dark matter.  The net result of these processes is the displacement of a significant portion of the surviving secondary core's material outwards to its tidal radius.  This material is then efficiently stripped away by tidal effects forming plume 'C' depicted in Fig. \ref{fig-3to1_plumes}b.  As a result, the plume formed from the secondary after apocentric passage assumes the morphology of a trailing tidal tail \citep{ToomreToomre72}.  More specifically, it has a shape which is curved in the same sense (from tail to core) as the orbit (see Fig. \ref{fig-3to1_plumes}b).  As the system passes apocentre, this structure precesses until it is oriented radially with the primary core.  Plumes ejected from primary cores at \tclosest\ are confined by the pressure gradients of the secondary's wake and do not share this shape nor evolution.

The symmetry of our 1:1 off-axis mergers means that all of the processes discussed above act on both cores equally to create a pair of symmetric plumes.  They connect at \taccrete\ forming an ``integral'' shaped structure.  Its curvature is much more pronounced in the $v_t/V_c=0.4$ case.  Comparing Figs. \ref{fig-1to1_entropy} to \ref{fig-3to1_entropy}, we can see that the plumes formed from the secondary cores are quite similar between the 1:1 and 3:1 cases.  However, in the 3:1 cases the plumes formed from the primary core reach much smaller radii.

The compressive forces generated in our 10:1 simulations are far less significant than in the 1:1 or 3:1 cases resulting in a much less significant plume of primary core material following \tclosest.  Furthermore, the resulting pressure gradients are far less localised and result in a broader and less collimated displacement of material.  By \tapo, all of the material of the secondary beyond $300$\kpc\ has been completely stripped and dispersed into the outer regions of the system.  Part of the core survives until \taccrete\ though, generating a small but undetectable plume after \tapo.

Several examples of observed systems exhibiting these morphologies exist including A1758 \citep{DavidKempner04} and A2744 \citep{KempnerDavid04}.  These systems possess disturbed accreting cores with curved shapes.  In both cases, the authors suggest that they are products of ram pressure stripping but this interpretation is not consistent with our simulations in either case: their large projected separations from their primary cores and their shape suggest they have ejected plumes similar to plume 'C'.  A1758N appears to have passed in front of A1758S from east to west, transited apocentric passage and is ejecting a plume of material as it returns back toward A1758S.  The A1758N plume is curved in the sense expected from such an orbit.  A plume from the core of A1758S is apparent but lacks the same curved appearance.  This may mean that A1758S is the more massive of the systems and has a plume such as plume ``B'' depicted in Fig. \ref{fig-3to1_plumes}a, or that projection effects are distorting its appearance.

A more secure example of a plume being ejected from a massive primary system such as plume ``B'' can be seen in A1644 \citep{Reiprichetal04}.  As noted above, this system has a significant bridge joining the interacting systems.  It appears to be at or near apocentric passage and the bridge's lack of curvature, the absence of curved plumes and the thickness of the bridge suggest that the system's orbit is highly inclined.  Although the temperature map is cut-off near the main system to the south west, there is a strong suggestion of an elongated cold region similar to plume ``B'' seen in a $y$-projection.   As we would expect from such an orientation, this feature is nearly collinear with the bridge.  

Another good example of a system which appears to have had a plume of primary core material ejected by a merger is A2744.  \citet{KempnerDavid04} suggest that this system is a 4:1 merger and should compare well to our 3:1 off-axis cases.  We conclude from the structure of the bridge joining the interacting systems and the shape of the secondary's plume that this system is near apocentric passage and significantly inclined.  The left-handed orientation of the secondary plume's curvature suggests that it has passed behind the primary in a north western direction and is now travelling towards us near apocentric passage.  Consistent with this interpretation we find a cold region to the south east of the primary, where the secondary would have entered the system, ejecting a plume from the primary's core.

\subsection{Induced core rotation}\label{analysis-rotation}

Due to the initial gas density gradients of the systems, the ram pressure acting on the cores as they pass each other is differential in off-axis cases.  This induces torques on the gaseous components of each system causing them to rotate.  Generally, the result is two co-rotating cores lasting from \tclosest\ to the second pericentric passage at \taccrete\ when the coherence of this motion is disturbed.  Remarkably, the release of the pressure gradients which lead to plume ``B'' in Fig. \ref{fig-3to1_plumes}a applies enough torque on the primary core in our 3:1 $v_t/V_c=0.15$ case to cause the core to counter-rotate.  This is the only instance in which we witness such counter rotation and more exploration of orbital and mass ratio parameter space is needed to understand how often this occurs.

As a result of its induced rotation prior to apocentric passage, our secondary cores have prograde rotations when they eject the material which form the tidal tail plumes discussed above.  This enhances the efficiency of the process.

\subsection{Streams}\label{analysis-streams}

\begin{figure}
\begin{center}
\leavevmode \epsfysize=8cm \epsfbox{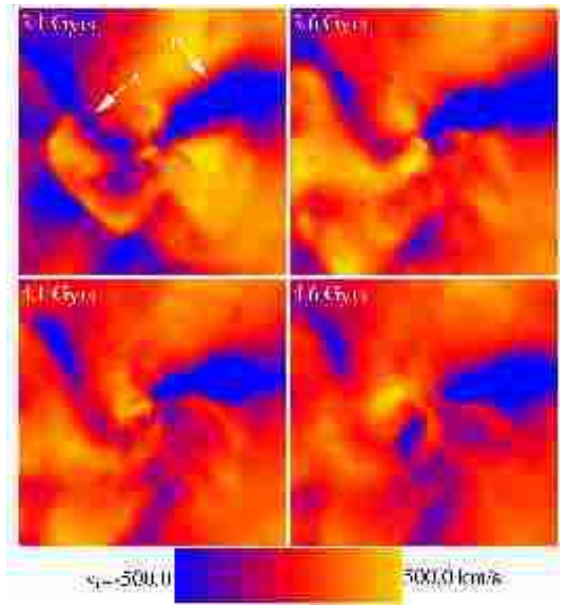}
\caption{Different time snapshots of the radial velocity maps (mass weighted) for the gaseous component in the 3:1 $v_t/V_c=0.15$ simulation
showing the collision of a high velocity stream (labelled B) with the remnant core.  In this specific case, the  primary and the secondary cores
merged at \taccrete$=2.5$\Gyrs.  Stream B consists of low entropy material that was originally part of the secondary core but shorn off during
the course of the merger.  Also shown is a stream feature (labelled A) that is not readily apparent in the surface brightness map but which
corresponds to a stream associated with returning material originally ejected from the primary core during the creation of plume `A' (see
Fig. \ref{fig-3to1_plumes}a).
Each frame here is $3.0$\Mpc\ on a side and is a $z$-projection of a $0.5$\Mpc\ thick slice$^\dagger$.}
\label{fig-3to1_vr_stream}
\end{center}
\end{figure}

The plumes of material removed from the cores of merging clusters discussed in Section \ref{analysis-plumes} collect into large collimated structures.  Because of the reduced entropy of this material, its buoyancy is low and it ultimately accretes onto the remnant core after the remainder of the secondary core does so at \taccrete.  As a result, long lived ($\sim 2$\Gyr) high velocity ($v \sim 1000$\kms) structures form in the cores of the remnant before the system relaxes.  We distinguish these structures from plumes because they are inwardly propagating high velocity structures of low (generally undetectable) surface brightness and are not associated with an apparent merging core (the secondary having been accreted and its observable presence destroyed at \taccrete).

In Fig. \ref{fig-3to1_vr_stream} we present a series of radial velocity maps for our 3:1 $v_t/V_c=0.15$ simulation illustrating the interaction of two low entropy streams (labelled ``A'' and ``B'') with the remnant core.  The secondary core accretes from the top right at \taccrete$=2.5$\Gyrs\ in this case and the stream formed from its disruption (labelled ``B'') follows it from the same direction.  Another inwardly radial (but unapparent in surface brightness) feature (labelled ``A'') can be seen on the opposite side of the system in these frames.  It is a long lasting stream formed from plume ``A'' identified in Fig. \ref{fig-3to1_plumes}a.

Large scale bulk motions with velocities similar to these streams were observed in the cores of merger remnants in the cosmological AMR simulations of \citet{Motl04}.  These authors raise concerns that such motions preclude the existence of actively cooling cores.  In our simulations, high velocity structures in the core remain highly collimated and completely disrupt cooling of the core in only one case (3:1 $v_t/V_c=0.15$).  Even in this extreme case, cooling resumes on rapid ($\sim 300$\Myrs) timescales once the stream finishes its accretion.  This takes $\sim2$\Gyrs.  In all other 1:1 and 3:1 off-axis cases we have studied, the high velocity streams possess enough angular momentum when they accrete to miss the region of active cooling.  In the 10:1 cases, the stream does not significantly disrupt cooling even when it does interact directly with the remnant core.  In \citet{P06c} we will examine in more detail the effect these streams can have on the final structure of a merger remnant.

The low entropy streams formed in the 1:1 off-axis cases are highly collimated and accrete quickly soon after \taccrete.  In all cases, some portion of the streams are visible until \taccrete\ when their appearance becomes disrupted by the second interaction of the cores.  In the $v_t/V_c=0.4$ case, the accreting streams form a large low entropy spiral which evolves into a ring roughly $1$\Mpc\ in diameter by \trelax\ (see Fig. \ref{fig-1to1_Tsl}).  This structure then slowly dissolves, disappearing $\sim2$\Gyrs\ after forming.

The secondary cores in our 3:1 mergers become more significantly disrupted after \tclosest\ than in the 1:1 cases.  As a result, material stripped from the secondary core becomes dispersed over a larger volume.  Furthermore, it accretes slightly later than \taccrete\ and over a longer period.  The streams are barely observable in our simulated \Chandra\ observations for only a $\sim 300$\Myrs\ following \taccrete.

Streams such as these have low surface brightness and are best observed in the temperature structure of clusters.  They should manifest as cold fronts oriented radially to the primary core with no obvious secondary core being present.  There are several examples of observed systems like this.  In A576 for instance, \citet{KempnerDavid04b} note a ``finger'' of cool gas which extends to the northwest of the system from its core.  In A2255 \citep{Sakelliouetal06} there is a distinct cool region to the southeast of the cluster core which is not associated with an obvious peak in X-ray brightness.  Finally, in A2034 \citep{Kempneretal03} there is a large diffuse tail of relatively cool gas extending towards the south of the cluster centre with no obvious X-ray peak associated with it.

\subsection{Edges}\label{analysis-edges}

The enhanced spatial resolution of \Chandra\ has revealed that the central regions of several clusters exhibit ``edges'' in their surface brightness as a result of large and localised gas density gradients.  Among these are RXJ1720.1+2638 \citep{Mazzottaetal01}, A496 \citep{DupkeWhite03} and A1795 \citep{Markevitchetal01}.  Systems with several of these features have also been found including A576 \citep{KempnerDavid04b} which has at least $2$ within $50$\kpc\ and ZW3146 \citep{Formanetal02} which has $3$ at $r\sim20$, $45$ \& $170$\kpc.  Attempts to account for these features have generally invoked bulk motions of cold core material with respect to its surrounding ICM.  Explored mechanisms capable of driving such motions include the activity of jets from central AGN \citep{Fabianetal05} or gravitationally induced motions due to small passing substructure \citep{Markevitchetal01,TittleyandHenriksen04}.

Occasionally when the streams discussed in Section \ref{analysis-streams} accrete to the centre of the remnant, an interface forms between it and the stationary core.  Significant and abrupt jumps in density, temperature and pressure can result accompanied by a jump in surface brightness reminiscent of observed edges.

\begin{figure}
\begin{center}
\leavevmode \epsfysize=8cm \epsfbox{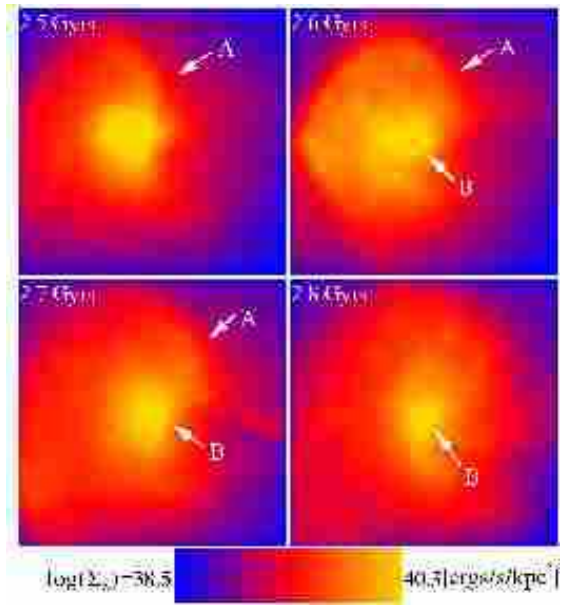}
\caption{Surface brightness maps of our 3:1 $v_t/V_c=0.15$ simulation showing the formation of $2$ ``edges'' associated with the accretion of the
secondary (labelled A) and the stream of low entropy material that follows the secondary (labelled B).  The first ``edge'' is weak feature that
moves outward in radius and disappears soon after \taccrete.  The second ``edge'' is more well-defined and remains stationary during the
accretion of the stream material, a process that can last for $\sim 0.5$\Gyrs.  We do not see any examples of ``edges'' produced by core
oscillations, nor do we find evidence for significant core oscillations in our simulations.
Each frame shown is a $z$-projection $0.5$\Mpc\ on a side$^\dagger$.}
\label{fig-3to1_edge}
\end{center}
\end{figure}

In Fig. \ref{fig-3to1_edge} we show the evolution of two such edges following \taccrete\ for our 3:1 $v_t/V_c=0.15$ simulation.  As the secondary core accretes to the primary core in this case, it does so on the top right side.  This leads to an expanding edge (labelled ``A'') which starts at a radius of $120$\kpc\ and disappears $0.3$\Gyrs\ later at a radius of $210$\kpc.  The stream which follows the accreting secondary core gets deflected towards the bottom of the core by this structure, forming a more significant edge (labelled ``B'') at a radius of $40$\kpc\ which lasts for $0.5$\Gyr.

In Section \ref{analysis-streams} we noted several systems which show evidence of streams of radially collimated material.  In two of these, A576 and A2034, there are obvious edge features present.  In the case of A576, these edges are aligned properly with the orientation of the stream to be a produced by it.  In the case of A2034, the edges are apparently on the opposite side from the stream.  In our simulations, multiple streams are typically produced by a single merger.  It may be that an unseen stream (possibly produced by the same event which generated the observed stream) is responsible for the observed edge in this case.  It may also be that the observed stream is passing in front of or behind the core before making contact with it.

No obvious edges are produced in our simulations by core oscillations.  Edges produced through the accretion of streams are distinct, long lived ($\sim 500$\Myrs) and can yield multiple edges (such as those seen in A576) simultaneously.  The last of these properties is difficult to account for through core oscillations but naturally explained as a product of accreting streams.


\section{Summary and conclusions}\label{sec-summary}
\begin{itemize}
\item This paper is the first in a series aimed at elucidating the impact of mergers on relaxed X-ray clusters.   For this study, we have carried out and analysed a suite of nine SPH simulations (incorporating the effects of cooling as well as star formation and associated SNe feedback) of two-body mergers between idealised clusters covering interesting and cosmologically relevant range of mass ratios and orbital properties.  Our merging clusters are initialised with realistic initial dark matter and gas distributions, the latter resembling those seen in relaxed clusters with cool compact cores observed by \Chandra\ and \XMM.  
\item Following a detailed qualitative analysis, we have identified a generic progression for typical cluster mergers.  The stages of this progression are: (i) pre-interaction, (ii) initial closest approach, (iii) apocentric passage, (iv) accretion and disruption of the secondary core and (v) relaxation.   To facilitate comparison and interpretation of observations, we describe in detail the interactions associated with the gas components of the two merging systems during each of the stages and quantify the times at which these stages arise relative to the beginning of the merger.  We also show gas density, X-ray surface brightness, gas temperature, SZ and gas entropy maps at different points during the mergers.

\item Moderate and massive mergers tend to run for $4.5$--$5.5$\Gyrs, while minor mergers last for up to  $\sim 2$\Gyrs~longer \citep[in good agreement with the trend identified by][]{Tormenetal04}.  This duration is the  elapsed time from the point when the secondary's centre of mass crosses $R_{200}$ of the primary (the beginning) to the time when the gas distribution in the final system has relaxed and nearly returned to virial/hydrostatic equilibrium (the end).  Typically, off-axis mergers run longer than head-on mergers.

\item In simulated $50 \ks$ \Chandra\ observations (assuming $z=0.1$), all easily identifiable substructure generated by our merger events disappear during the secondary's second pericentric passage of the primary core.  However, the system takes an additional $\sim 2$\Gyrs~for its isophots to appear relaxed.  Temperature fluctuations are not necessarily a reliable indicator of the system's dynamical state.

\item We find that following the merger, the resultant system settles into virial equilibrium sooner than into hydrostatic equilibrium.  In many cases, the final cluster exhibits oscillations that result in instantaneous deviations from hydrostatic equilibrium by as much as $15$\% for as long as $7$ Gyrs after the beginning of the merger.  We also find that generically complex patterns of entropy and temperature fluctuations, at the level of  $\Delta T/T\sim 20\%$, can persist in the final systems well past the point when the systems appears relaxed.   Based on our findings, we suggest that temperature fluctuations of this order, without any other corroborating signatures of ongoing merger, should not be interpreted as evidence of a highly disturbed system.  

\item We find that in none of the cases considered are the initial cool compact cores of the primary and the secondary destroyed during the course of the mergers.  Instead, the two remnant cores eventually combine to form a new core that, depending on the final mass of the remnant, can have a greater cooling efficiency than either of its progenitors. We will discuss this behaviour in more detail in a forthcoming paper.

\item While the large scale coherent displacements between the dark matter and gas components in the merging system are not uncommon, they do not trigger any
obvious surface brightness discontinuities.   Within the central $50$\kpc, the gas and the dark matter remain tightly coupled to each other.   We do not see any
evidence of core oscillations or of any associated heating.

\item We investigated the efficacy of three measures used to quantify the degree to which observed clusters are disturbed.  These are (i) centroid variance, (ii)
power ratios, and (iii) X-ray surface brightness/projected mass displacement.  We find that the centroid variance is the best of the three.  It best reflects the state of the cluster as determined by visual examination.  It also provides an excellent indicator of how far the system is from virial and
hydrostatic equilibrium.  The power ratios are very sensitive to noise and once this is factored in, they provide an unambiguous signal only when the system is
highly disturbed.  The X-ray/mass displacement are extremely sensitive to weak but long-lived perturbations in the dark matter distribution caused by low mass
mergers.

\item During the course of the merger, we find a variety of transient features arising and disappearing in the projected temperature, entropy and surface
brightness maps, most of which resemble ``cold fronts'' observed in recent \Chandra\ and \XMM\ observations.  While ``cold fronts'' are generally associated with
the disruption of the secondary's core, we find that they can be caused by several different mechanisms.  The resulting features can potentially be distinguished
based on their morphological properties.  Interested readers are referred to Section \ref{analysis-transients} for examples of clusters whose recently published
X-ray images show examples of the different types of transient structures.  We propose the following classification scheme based on their origin and appearance:

\begin{itemize}
\item \emph{``Comet-like'' tails}: Ram pressure strips the outer atmospheres of secondary systems forming very brief ($\sim 0.5$\Gyr) comet-like morphologies for both cluster cores (except the primary in 10:1 mergers).  This morphology is observable only between first pericentric and apocentric passages.
\item \emph{Bridges}: Ram pressure disperses core material from both cluster cores in 1:1 and 3:1 cases.  A luminous moderate-entropy bridge resulting from the convergence of the two resulting regions of dispersed material is produced.  This structure remains observable in $50$\ks\ \Chandra\ exposures at $z=0.1$ until secondary core accretion in our 1:1 mergers and until apocentric passage in our 3:1 mergers.  The morphology of this structure could be useful as a cue for determining the inclination of observed merger systems.
\item \emph{Plumes}: Pressure gradients produced by the secondary's motion at \tclosest\ and \tapo\ can lead to the ejection of significant amounts of the primary and secondary cores' gas.  This material forms large collimated plumes of low-entropy material.  At apocentric passage, material ejected in this way from the secondary's core becomes tidally stripped and adopts the morphology of a trailing tidal tail.  The majority of observed disturbed merging cores likely owe their appearance to this process, rather than ram pressure stripping.
\item \emph{Streams}: Low entropy streams form from these plumes and accrete to the centre of the remnant following the second pericentric passage and disruption of the secondary core at \taccrete.  This material remains collimated and typically has enough angular momentum to avoid a direct impact with the remnant core.  This mitigates the concerns of \citet{Motl04} who suggest that high velocity structures formed in mergers may prevent the formation of compact cool cores.  These streams may also play an important role in fuelling the activity of central AGN in clusters.
\item \emph{Induced core rotation}: Off-axis mergers subject the gaseous cores of both merging clusters to torques which induce rotation.  In one simulation (our 3:1 $v_t/V_c=0.15$ case), the resulting rotation of the primary's core is retrograde to the secondary's orbit.  In all cases, the coherence of this rotation is completely disrupted by the disruption of the cores during second pericentric passage.
\item \emph{Edges}:  Streams can accrete directly onto the core of the remnant, generating strong ``edges'' in surface brightness and gradients in temperature and density.  The resulting features are similar to several cold front systems whose morphology had previously been attributed to AGN activity or motion of the gaseous core within its dark matter potential.
\end{itemize}
\end{itemize}

We would like to reiterate that our simulations do not incorporate the physical processes associated with magnetic fields, pressure from cosmic rays or thermal conduction nor do we adequately model the effects of turbulence.  The influence of these processes on clusters remain contentious issues and we can not make any confident predictions at this point regarding their effects on our results.

\section*{Acknowledgements}
We would like to thank Josh Barnes for allowing us use of his ZENO software package.  We are also grateful to Michael Balogh, Andi Mahdavi, Joe Mohr, Scott Kay, Graham Smith and Megan Donahue for stimulating discussions and insightful comments.  IGM acknowledges support from a NSERC Postdoctoral Fellowship and a PPARC rolling grant for extragalactic astronomy and cosmology at the University of Durham.  AB acknowledges support from NSERC through the Discovery Grant program.  MF acknowledges support from NSERC, NASA, and NSF.


\label{lastpage}

\end{document}